\documentclass{article}
\usepackage{arxiv}

\usepackage[utf8]{inputenc} 
\usepackage[T1]{fontenc}    
\usepackage{hyperref}       
\usepackage{url}            
\usepackage{booktabs}       
\usepackage{amsfonts}       
\usepackage{nicefrac}       
\usepackage{microtype}      
\usepackage{lipsum}		
\usepackage{graphicx}
\usepackage[numbers]{natbib}  

\usepackage{doi}
\usepackage{subcaption} 
\usepackage{color} 
\usepackage{adjustbox}
\usepackage{amsmath}
\usepackage{mathtools, nccmath}
\usepackage{xparse}
\usepackage{gensymb} 
\newcommand{\beginsupplement}{%
        \setcounter{table}{0}
        \renewcommand{\thetable}{S\arabic{table}}%
        \setcounter{figure}{0}
        \renewcommand{\thefigure}{S\arabic{figure}}%
     }

\usepackage{algorithm}
\usepackage{algpseudocode}
\usepackage{bm}
\usepackage{multirow}
\usepackage{lscape}
\usepackage{longtable}
\usepackage{textcomp}

\title{Efficient proton arc optimization and delivery through energy layer pre-selection and post-filtering}

\usepackage{authblk}

\author[1,*]{Sophie Wuyckens}
\author[2]{Viktor Wase}
\author[2]{Otte Marthin}
\author[2]{Johan Sundstr\"om}
\author[3,4]{Guillaume Janssens}
\author[1]{Elena Borderias-Villarroel}
\author[1,4]{Kevin Souris}
\author[1,5,6]{Edmond Sterpin}
\author[2]{Erik Engwall}
\author[1]{John A. Lee}

\affil[1]{UCLouvain, Institut de recherche expérimentale et clinique, Molecular Imaging and Radiation Oncology Laboratory, Brussels, Belgium}
\affil[2]{RaySearch Laboratories, Stockholm, Sweden}
\affil[3]{Ion Beam Applications SA, Louvain-La-Neuve 1348, Belgium}
\affil[4]{UCLouvain, Institute of Information and Communication Technologies, Louvain-La-Neuve, Belgium}
\affil[5]{KULeuven, Department of Oncology, Laboratory of experimental radiotherapy, Leuven, Belgium}
\affil[6]{Particle Therapy Interuniversity Center Leuven - PARTICLE, Leuven, Belgium}
\affil[*]{Corresponding author: Sophie Wuyckens, sophie.wuyckens@uclouvain.be}

\renewcommand{\headeright}{Preprint}
\renewcommand{\undertitle}{Preprint}
\renewcommand{\shorttitle}{Optimizing proton arc efficiency}

\hypersetup{
pdftitle={Optimizing proton arc efficiency},
pdfsubject={},
pdfauthor={S.~Wuyckens, V.~Wase, O.~Marthin J.~Johan Sundstr\"om, G.~Janssens, E.~Borderias-Villarroel, K.~Souris, E.~Sterpin, E.~Engwall, J. A.~Lee},
pdfkeywords={optimization, proton arc therapy, treatment planning},
}

\begin{document}

\maketitle

\begin{abstract}
\noindent {\bf Background:} Proton arc therapy (PAT) has emerged as a promising approach for improving dose distribution, but also enabling simpler and faster treatment delivery in comparison to conventional proton treatments. However, the delivery speed achievable in proton arc relies on dedicated algorithms, which currently do not generate plans with a clear speed-up and sometimes even result in increased delivery time. \\ 
{\bf Purpose:} This study aims to address the challenge of minimizing delivery time through a hybrid method combining a fast geometry-based energy layer (EL) pre-selection with a dose-based EL filtering, and comparing its performance to a baseline approach without filtering.\\
{\bf Methods:} Three methods of EL filtering were developed: unrestricted, switch-up (SU), and switch-up gap (SU gap) filtering. The unrestricted method filters the lowest weighted EL while the SU gap filtering removes the EL around a new SU to minimize the gantry rotation braking. The SU filtering removes the lowest weighted group of EL that includes a SU. These filters were combined with the RayStation dynamic proton arc optimization framework (ELSA). Four bilateral oropharyngeal and four lung cancer patients' data were used for evaluation. Objective function values, target coverage robustness, organ-at-risk doses and NTCP evaluations, as well as comparisons to IMPT plans, were used to assess plan quality. \\
{\bf Results:} The SU gap filtering algorithm performed best in five out of the eight cases, maintaining plan quality within tolerance while reducing beam delivery time, in particular for the oropharyngeal cohort. It achieved up to approximately 22\% and 15\% reduction in delivery time for oropharyngeal and lung treatment sites, respectively. The unrestricted filtering algorithm followed closely. In contrast, the SU filtering showed limited improvement, suppressing one or two SU without substantial delivery time shortening. Robust target coverage was kept within 1\% of variation compared to the PAT baseline plan while organs-at-risk doses slightly decreased or kept about the same for all patients.  \\
{\bf Conclusions:} This study provides insights to accelerate PAT delivery without compromising plan quality. These advancements could enhance treatment efficiency and patient throughput. \\

\end{abstract}

\newpage     


\newpage

\setlength{\baselineskip}{0.7cm}      

\pagenumbering{arabic}
\setcounter{page}{1}
\pagestyle{fancy}
\section{Introduction}
Over the past decade, proton arc therapy (PAT) has emerged as an alternative delivery mode to fixed gantry intensity-modulated proton therapy (IMPT). This treatment technique involves continuous rotation of the gantry while delivering radiation. As such, it allows for more freedom to administer the proton treatment from many angles, similar to Volumetric Modulated Arc Therapy (VMAT) in photon therapy \cite{otto_volumetric_2007}.

Two publications in the mid-1990's initiated proton arc research: one conceptualized ``proton tomotherapy'' treatments\cite{deasy_conformal_1995}, while the other explored the feasibility of proton arc using passive scattering (PS) with a rotating anthropomorphic phantom \cite{sandison_phantom_1997}. However, 15 years passed before proton arc resurfaced, largely due to the advent of proton pencil beam scanning (PBS) \cite{pedroni_200-mev_1995}. PBS delivery addressed several technical challenges that PS could not, such as the difficulty of adjusting range modulation wheel settings during simultaneous gantry rotation and irradiation \cite{seco_proton_2013}. Following that breakthrough, several studies across various disease sites were published and  aimed at demonstrating the potential of proton arc for improved treatment outcomes \cite{ding_have_2018,ding_improving_2019,li_improve_2018,liu_improve_2020,liu_novel_2020,de_jong_proton_2023}. Concurrently, researchers have been developing treatment planning optimization algorithms to effectively utilize the new degrees of freedom brought by the many field directions and spot candidates in arc delivery, while also ensuring reasonable treatment delivery time \cite{ding_spot-scanning_2016,battinelli_proton_2019,engwall_fast_2022,zhang_energy_2022,wuyckens_treatment_2022}.

One of the anticipated advantages of PAT is the potential for faster treatment delivery compared to conventional proton treatments as was seen with VMAT when compared to fixed gantry intensity-modulated radiotherapy \cite{teoh_volumetric_2011}. Several proton arc studies have already demonstrated that PAT can reduce overall treatment time \cite{chang_feasibility_2020,liu_improve_2020,ding_improving_2019,li_improve_2018}. Importantly  though, reported delivery times are often estimated with rough approximations, relying solely on the total energy layer switching time (ELST), without accounting for the distinction between upward and downward energy switching, likely due to the limitations of different proton machines available in the market \cite{van_de_water_shortening_2020,krieger_quantitative_2022}. In the proton arc studies cited above, the arc timing advantage was actually only significant when ELST fell below 0.5 s, but this assumption for upward energy switching is not realistic in most of the currently available hardware. This makes it challenging to draw definitive conclusions based on these estimates only. Furthermore, contradictory findings have also been reported, where PAT exhibited prolonged treatment delivery durations compared to IMPT \cite{ding_have_2018,wera_proton_2023}, especially in the discrete arc mode where the delivery entails a large number of fixed-angle fields for which the gantry needs to come to a full stop at each discrete direction to deliver a number of stacked energy layers \cite{de_jong_spot_2023}. In the dynamic mode, the prolonged delivery times stem from a combination of several factors, including energy layer switching time, spot irradiation, gantry rotation, a large number of layers and spots, and the difficulty of accurately delivering radiation from the correct directions within a tolerance window, while rotating the gantry. Consequently, the gantry needs to undergo frequent deceleration and acceleration \cite{wase_optimizing_2024}, e.g., decelerate to allow buffer time for energy switching before delivering the next energy layer or accelerate when some angles are not used for irradiation. 

Several independent research groups have therefore dedicated efforts to devise approaches to minimize the delivery time of proton arc plans. One such approach is to formulate additional objectives to be minimized, which are surrogate measures of the delivery time, while simultaneously optimizing the dose distribution within an integrated framework \cite{gu_novel_2020,zhang_energy_2022,wuyckens_treatment_2022,zhao_first_2023}. Since dose objectives, together with the delivery time objectives, influence the energy layer selection, it is referred to as a \textit{dose-based} method in this study. While the concept is elegant in theory, this approach encounters practical and mathematical challenges \cite{wase_proton_2024}. These include the requirement for simplified mathematical models to estimate beam delivery time, the need for large computational resources and suitable solvers, as well as the necessity for careful tuning of objective weights. These limitations hinder its practical use and effectiveness in achieving substantial time savings in PAT planning and delivery. Another approach involves the selection of a user-defined number of energy layers using heuristics based on dosimetry, i.e., steered by the dose objectives during the spot weight optimization, making it also a dose-based approach. For instance, the spot-scanning proton arc (SPArc) therapy algorithm \cite{ding_spot-scanning_2016,liu_novel_2020} has already demonstrated efficacy across various disease sites. Its main drawbacks are long optimization times and limited flexibility to select energy switch locations. The pre-selection of the energy layers can also be done independently, before the spot optimization happens, but then only based on geometrical considerations, such as tumor coverage computation in terms of number of voxels covered by a specific sequence of energy layers \cite{engwall_fast_2022,cao_intensity_2023}. These \textit{geometrical-based} approaches include heuristics to control the maximum number of upward energy switches in the solution. For example, the generation of dynamic arc plans with early energy layer selection and spot assignment \cite{engwall_fast_2022} (ELSA) prior to robust spot weight optimization has been noticed for its fast execution time. This is due to two main factors: firstly, it reduces the spot weight optimization problem size, and secondly, it reduces the need to compute dose for a vast number of potential energy layers and spots. However, using these fast methods comes at a cost, as it constrains the optimizer freedom for the subsequent dose optimization process. 

To bridge the gap between the geometry-based and dose-based approaches, we propose the first hybrid method that combines their mutual advantages; the fast energy layer pre-selection combined with energy layer post-selection based on dose objectives, namely the energy layer filtering. In this study, the hybrid approach extends ELSA that rapidly creates an initial set of energy layers, which are filtered after robust spot weight optimization to a final set using information from the optimized dose distribution. As a result, the filtered plan accelerates delivery since there are fewer energy layers or spots to be delivered. This approach is inspired by the iterative filtering step in the SPArc algorithm 
\cite{ding_spot-scanning_2016} and the Reverse Greedy algorithm \cite{battinelli_proton_2019}, the latter being proposed for discrete proton arcs. These algorithms progressively eliminate the energy layers contributing the least to the plan quality. However, they do it concurrently during the spot weight optimization, and thus, the optimization efficiency is negatively impacted. We have developed and compared three distinct methods capable of energy layer filtering: The \textit{unrestricted filtering} that removes the lowest MU-weighted energy layers, the \textit{SU gap filtering} that filters energy layers to minimize the gantry deceleration before switching the energy upward and finally the \textit{SU filtering} that filters the lowest MU-weighted sequence of energy layers that includes at least one energy switch-up. These methods act therefore as post-processing tools that filter once the spot weight optimization process is completed. Through the implementation of these filtering techniques, the aim of the study is to enhance the overall efficiency of PAT delivery.

\section{Material and methods}
\subsection{Energy layer filtering algorithms}
\label{subsec6:algo}


Unlike IMPT plans, where each beam angle is optimized to cover the target in depth using a decreasing energy sequence, proton arc plans typically adopt a sector-based approach. Arc beams are divided into sectors, with energy decreasing within each sector and increasing only between sectors. This design stems from the anticipated increased time required for upward energy switches. Delivery occurs within fixed intervals (centered around a discrete direction) known as control points across each sector. 

The number of sectors in a proton arc plan, which defines the number of switch-ups (SU) between energy levels, is a carefully studied topic as it directly correlates with the delivery time for most machines \cite{gu_novel_2020,wuyckens_treatment_2022,zhang_energy_2022,zhang_treatment_2023}. The basic assumption of this conceptual model is that delivery time is proportional to the irradiation time of each energy layer and the number of switch-ups in the plan. Therefore, minimizing the delivery time can be achieved by filtering energy layers and/or reducing the number of SU.

In this study, methods are implemented and proton arc plans are optimized in a modified research build of the RayStation dynamic proton arc optimizer. While not exactly the same as the one showcased in the original publication \cite{engwall_fast_2022}, our approach still follows the ELSA framework for selecting energy layers based on geometry followed by the classical spot weight optimization. 

\begin{figure}%
    \centering
    \subfloat[\centering Baseline]{{\includegraphics[width=0.5\textwidth]{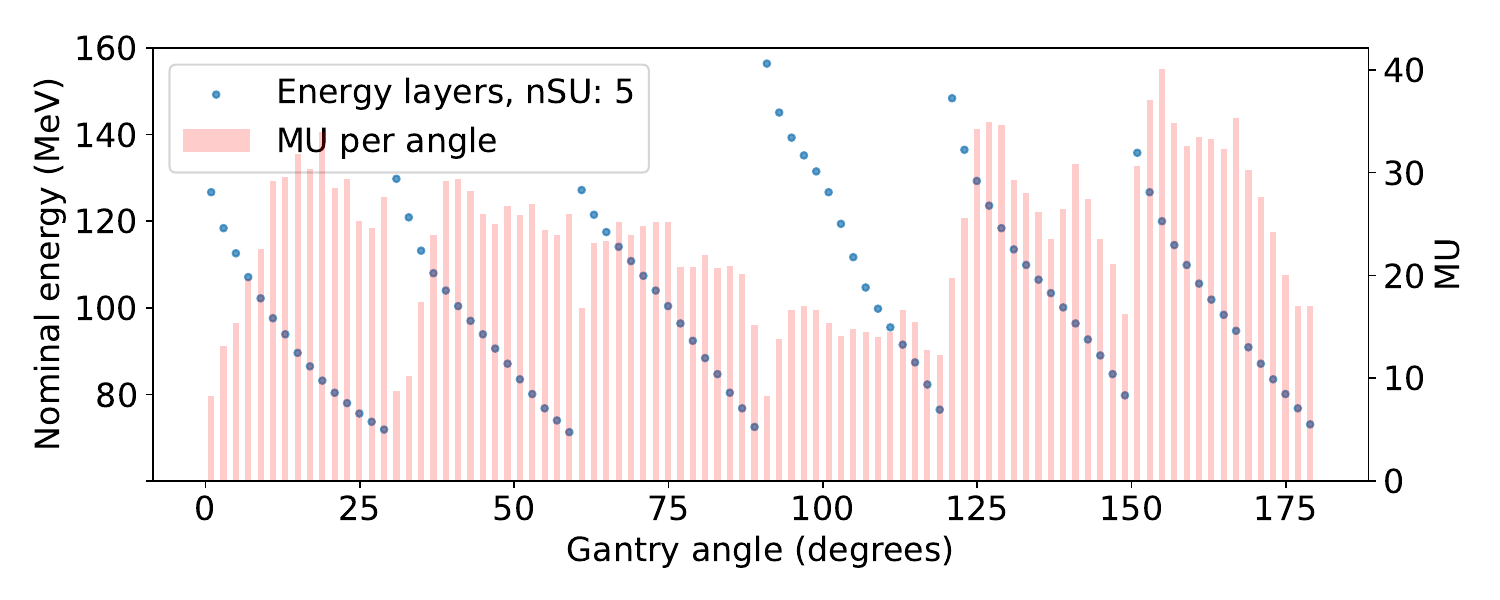} }}%
    \subfloat[\centering Unrestricted filtering]{{\includegraphics[width=0.5\textwidth]{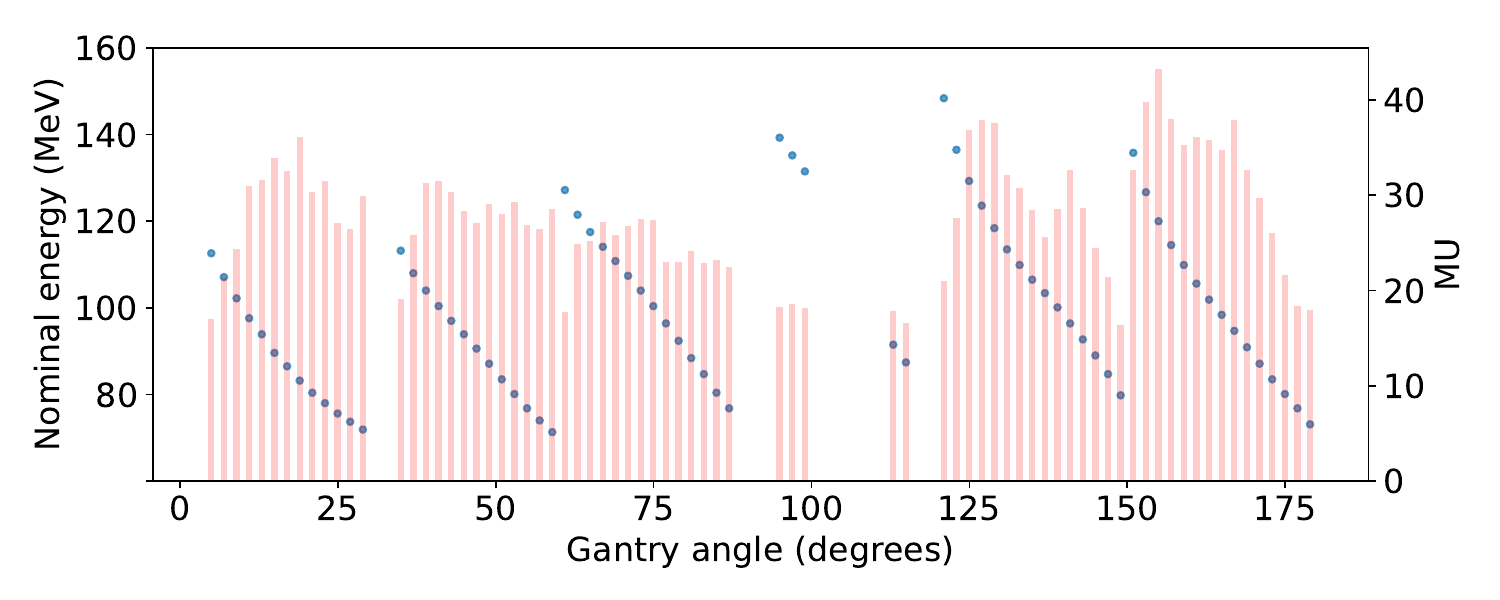} }}%
    \qquad
    \subfloat[\centering SU filtering]{{\includegraphics[width=0.5\textwidth]{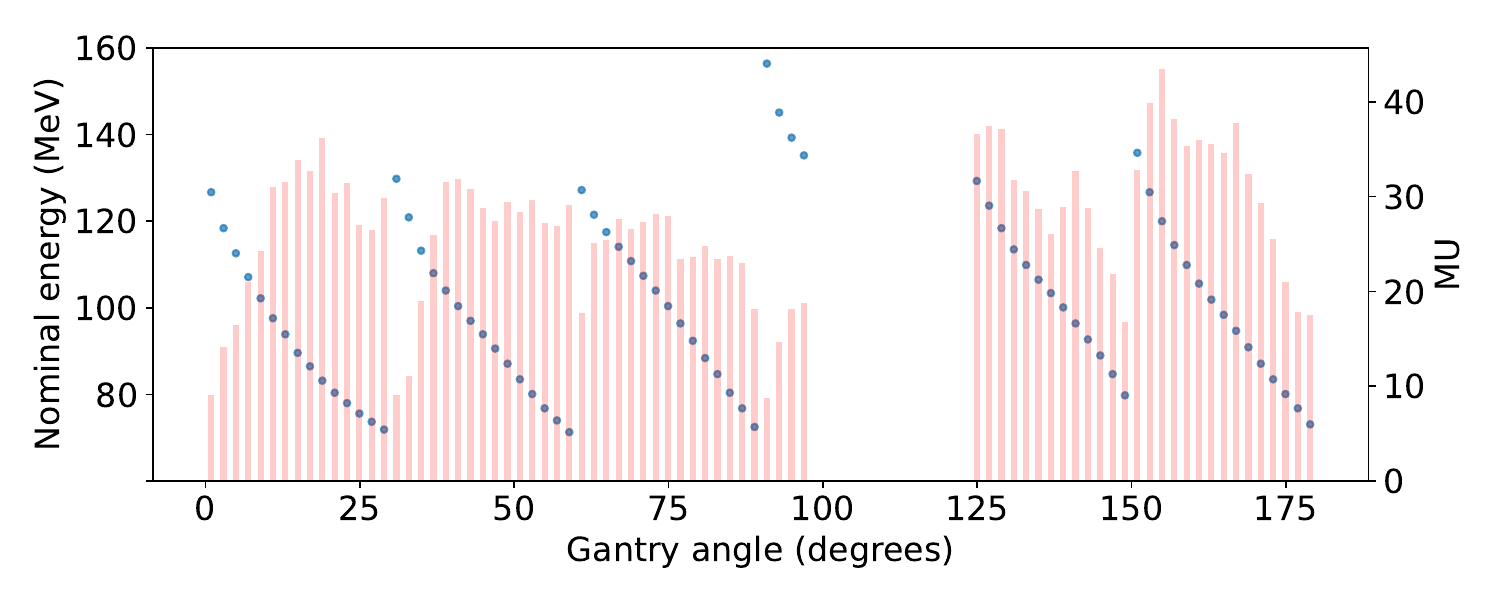} }}%
    \subfloat[\centering SU gap filtering]{{\includegraphics[width=0.5\textwidth]{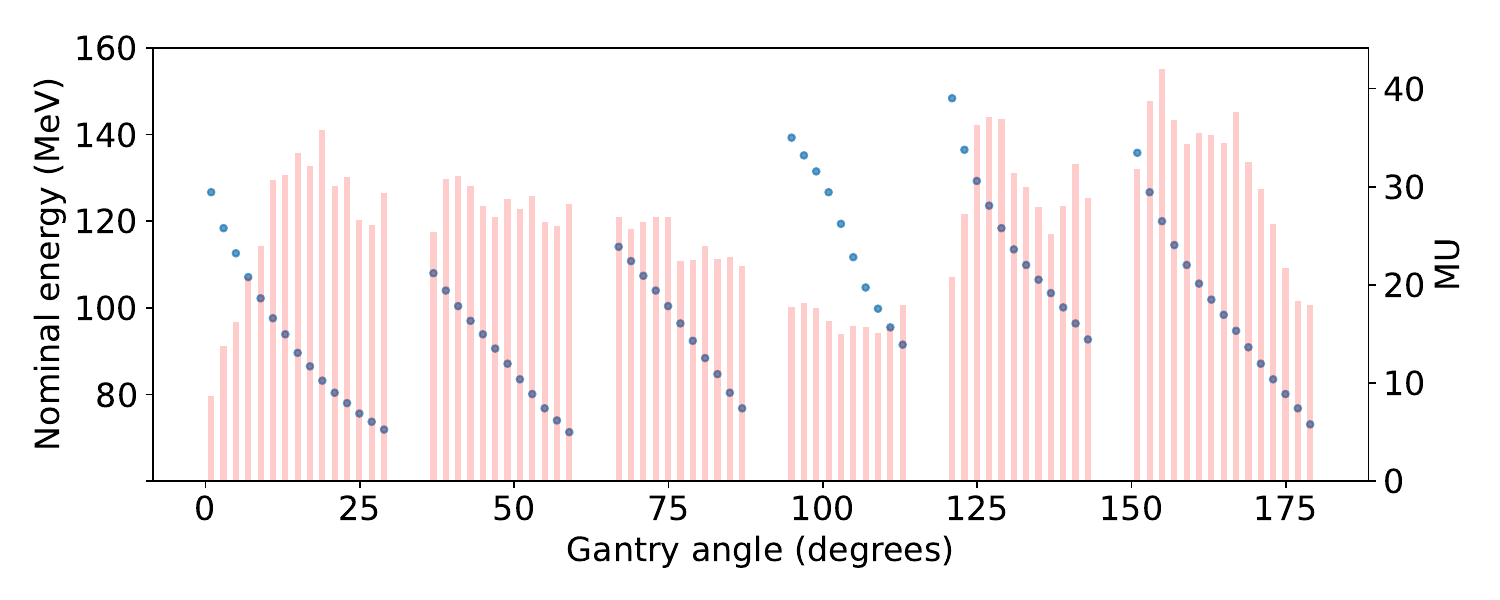} }}%
    \caption{Strategies to filter energy layers (EL) illustrated by a part of an arc for an example case. (a) Nominal arc plan used as starting point. (b) Unrestricted filtering removes the 15 lowest MU (EL). (c) SU gap filters the 15 EL that reduce the gantry time to break before each SU. (d) SU filters the lowest MU partial sector that includes at least one SU. }%
    \label{fig6:teffAlgo}%
\end{figure}

Considering an existing arc plan for an example case as a starting point (Fig.~\ref{fig6:teffAlgo}A), three distinct methods were developed to filter energy layers (EL) and/or reduce the number of SU as a post-processing step to the main spot weight optimization process. Assuming $N$ initial energy layers and $M$ initial SU in the plan, the goal is to keep at most $X$ energy layers or $Y$ SU. The three methods are summarized below:

\begin{enumerate}
    \item \textbf{Unrestricted filtering}: This algorithm filters the $N-X$ energy layers with the lowest corresponding monitor unit (MU)  from the plan. By removing the lowest MU energy layers, the aim is to reduce the overall number of energy layers in the plan while preserving the important dose contributions, i.e., the energy layers that accumulates the higher spot weights in total. In Fig.~\ref{fig6:teffAlgo}, from (a) to (b) individual holes appear where  the lowest EL have been suppressed.
    \label{method6:1}
    
    \item \textbf{SU filtering}: This algorithm filters $M-Y$ SU by deactivating the partial sectors with the lowest corresponding MU. It iterates through two nested loops over the energy layers in the plan. It checks if at least one SU occurs between the two indexed energy layers and verifies that the energy of the second indexed angle is lower than the energy of the first indexed angle. When these conditions are met, the algorithm accumulates the MUs of the energy layers between the two indexes, forming a partial sector. These partial sectors, containing SU, are then ordered based on increasing MU values. The energy layers included in each partial sector are subsequently removed from the plan. By eliminating the energy layers associated with partial sectors containing SU and having the lowest MUs, the SU filter effectively reduces the overall number of SU and EL in the plan. The SU-filtered plan in Fig.~\ref{fig6:teffAlgo}C effectively exhibits 4 energy SU, compared to 5 SU as is seen in the other three proton arc filtering techniques.
    \label{method6:2}  

    \item \textbf{SU gap filtering}: This algorithm removes the energy layers with the lowest MU, similarly to the unrestricted filtering algorithm. The difference is that this algorithm is restricted to only choose from the energy layers that are located just before or just after a SU. Note that after an energy layer is removed to create a gap, the adjacent energy layers on either side of the gap become the ones positioned just before and just after a SU.  The rationale behind this approach is that the gantry is often forced to decelerate to a rather low velocity in conjunction with a SU and increasing the corresponding angular span might mitigate this effect. The algorithm contains a maximum size of this gap, since at a certain point the gantry will be able to keep a constant speed and still perform a SU. This maximum gap size, measured in terms of the number of removed layers, is established as follows:

    \begin{equation}
        \left\lfloor\frac{t_{\text{SU}} + t^{\text{mean}}_{\text{EL}}}{t_{\text{SD}} + t^{\text{mean}}_{\text{EL}}} \right\rfloor
    \label{eq6:SUGap}
    \end{equation}

    where $\left\lfloor \cdot \right\rfloor$ is the floor function, $t_{\text{SU}}$ and $t_{\text{SD}}$ are the times to switch the energy upward and downward respectively and $t^{\text{mean}}_{\text{EL}}$ is the average irradiation time per energy layer. Which is to say, this is the ratio between the delivery time of a layer with an SU and a layer with a switch-down (SD). Note that this means that different plans may have different maximum gap sizes.

    \label{method6:3}

\end{enumerate}

Each algorithm, depicted in Fig.~\ref{fig6:teffAlgo}, is run multiple times, decreasing monotonically the allotted budget of energy layers $X$ (for unrestricted and SU gap filters) or switch-ups $Y$ (for SU filtering) in the plan, the so-called $\text{MaxBoundSequence}$. Following each filtering operation, the spot weight optimization is re-run with a reduced number of iterations to compensate for the loss of monitor units (MU) and to maintain plan quality. Each filtering run is independent from the others and start from the initial optimized spot weights of the optimized arc plan.

This strategy allows to populate a Pareto front for each algorithm tested. A Pareto front represents the set of optimal solutions in a multi-objective optimization problem, where improving one objective comes at the expense of worsening another \cite{craft_exploration_2005,wuyckens_bi-criteria_2022,fu_simultaneous_2024}. In this context, the Pareto front generated relates the attainable dosimetric objective values to the necessary dynamic delivery times. By examining the trade-off between these two sides of the problem, one can select the most suitable trade-off plan from the Pareto front. A pseudo-code summarizing the filtering workflow algorithm is provided in SM1.   

\subsection{Patient data and treatment planning}
\label{subsec6:patient}

To demonstrate the efficiency of the algorithms, two small cohorts of patients were selected: four patients with oropharyngeal cancer (OPC) and four patients with unresectable lung cancer. The patients all come from anonymized databases obtained from Cliniques Universitaires Saint-Luc in Brussels, Belgium. The local ethics committee approved the retrospective use of these databases. 



The clinical plans for both databases were all planned and delivered with photon modality. Nevertheless, the databases also includes IMPT plans, used for this study, which were manually generated by experienced dosimetrists. All information regarding patient IMPT planning protocol for both disease sites can be found in Tables S1, S2 and S3. The IMPT treatments were planned retrospectively in the treatment planning system (TPS) RayStation 11B (Raysearch Laboratories, AB), while all proton arc plans, referred to as PAT plans were planned in a research version of RayStation 2023B. Filtered PAT plans, employed the same set of clinical goals and robust settings as the original PAT plans which were initially obtained from the IMPT plans. However, in order to expedite the planning process, the total number of iterations for optimization was reduced to 50. All plans were normalized to the median dose (D50).

Following Algorithm S1, given $R=10$ runs, the $\text{MaxBoundSequence}$ was set to be 10 evenly spaced numbers over different intervals depending on the patient. For the whole OPC cohort, given the 360 initial EL and the 16 initial SU in each PAT plan, the sequence used for the maximum number of allowed energy layers (Method \ref{method6:1} \& \ref{method6:3}) is $\left[350,200\right]$ and the sequence for the maximum number of allowed SU (Method \ref{method6:2}) is a list of numbers decremented by 1 over the interval $\left[15,5\right]$ . These specific numbers were chosen intentionally to ensure a smooth transition towards the number of energy layers present in the IMPT plans, which is approximately 160 energy layers. For the lung cohort, each sequence was tailored to each patient depending on the initial number of energy layers.

\subsection{Treatment plan evaluation}

\subsubsection{Delivery time}

Since the objective of this study is to achieve faster delivery rather than improving plan quality, it is noteworthy that suppressing energy layers from the initial set of energy layers is not expected to enhance plan quality. The aim is merely to minimize the delivery time with precautions to ensure that plan quality remains comparable to the PAT baseline plan.

Dynamic beam delivery time (BDT) was calculated using the ATOM (Arc Trajectory Optimization Method) algorithm, an open-source tool \cite{wase_optimizing_2024} published by RaySearch Laboratories. ATOM is designed to determine a fast plan delivery while adhering to the mechanical constraints specified by the user. The machine delivery assumptions utilized in this study are detailed in Table S4.  The selection of ATOM was driven by its utility, given that no proton arc machine has yet been established and commissioned in clinical settings; ATOM serves as a reliable estimator at present. Nevertheless, in the final evaluation, BDT estimations obtained with ATOM were also provided but using the layer irradiation times and layer switching times based on a simulation of an IBA ProteusPlus machine using the IBA ScanAlgo system.

\subsubsection{Plan quality}

In addition to the primary objective of achieving faster delivery, a comprehensive dosimetric evaluation was included to consider the selected trade-off from the generated Pareto front. Target coverage was evaluated in the IMPT, PAT, and filtered PAT plans employing the worst-case scenario robustness evaluation approach in both cohorts. For OPC patients, 14 different isocenter shifts were considered, comprising 6 points along the main axes and 8 points along the diagonals, which are defined by the setup uncertainty. Moreover, three density shifts were taken into account, resulting in a total of 42 scenarios. The assessment of robust target coverage was based on the criterion of achieving a minimum dose of at least 95\% of the prescribed dose (Dp) at 98\% of the target volumes in a worst-case scenario dose distribution (D98$\%\geq95$\% Dp). This criterion was applied to both the high-risk and prophylactic lymph nodal regions, ensuring sufficient dose coverage for these target areas. Both arc plans (PAT and filtered PAT) were compared by evaluating the deviations in target coverage from the values achieved by the IMPT plan in the nominal and worst-case scenarios. The nominal average doses received by organs at risk in the head and neck region were also assessed using the same comparative approach. Several dose metrics on organs at risk (OARs) were also evaluated in the nominal plan and over the robustness evaluation scenarios computing the means and standard deviations. Finally, the evaluation of NTCP for xerostomia and dysphagia using the models from the Dutch model-based approach \cite{langendijk_indicatie_2019} was incorporated. The $\Delta$NTCP was calculated by taking the difference between the arc plans and the IMPT plan in the nominal scenario for each patient. For the lung cohort, the robustness evaluation considered 6 isocenter shifts along the main axes and, 3 density shifts and for 4 breathing phases, resulting in a total of 84 scenarios. Robust CTV coverage was assessed by evaluating D95\% and D98\%, in both the nominal and worst-case scenarios. Additional dose metrics were also reported to assess OAR robustness.

\section{Results}
\subsection{Energy layer filtering algorithms comparison}

\begin{figure}[!ht]%
    \centering
    \subfloat[\centering OPC 1]{{\includegraphics[width=8cm]{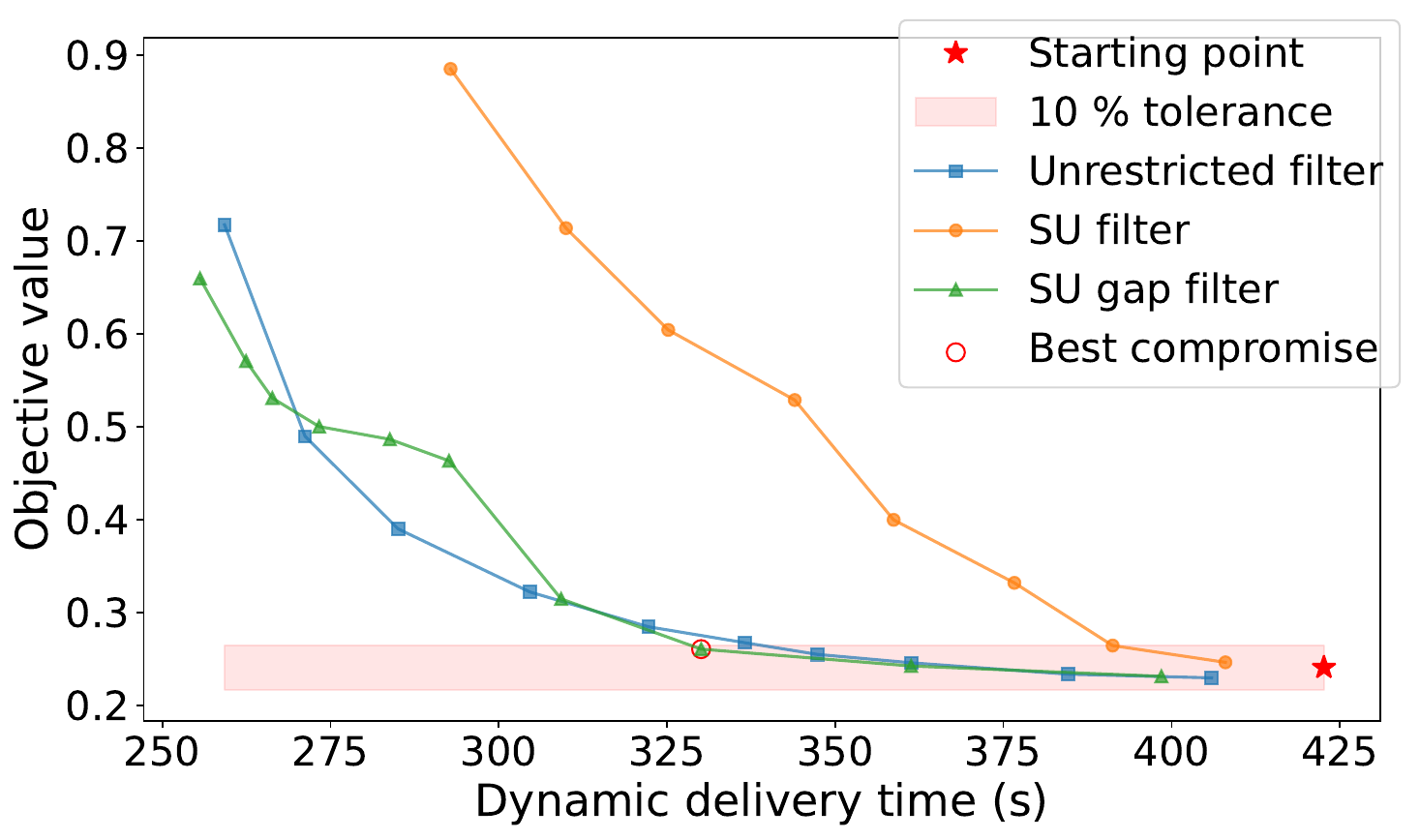} }}%
    \subfloat[\centering OPC 2]{{\includegraphics[width=8cm]{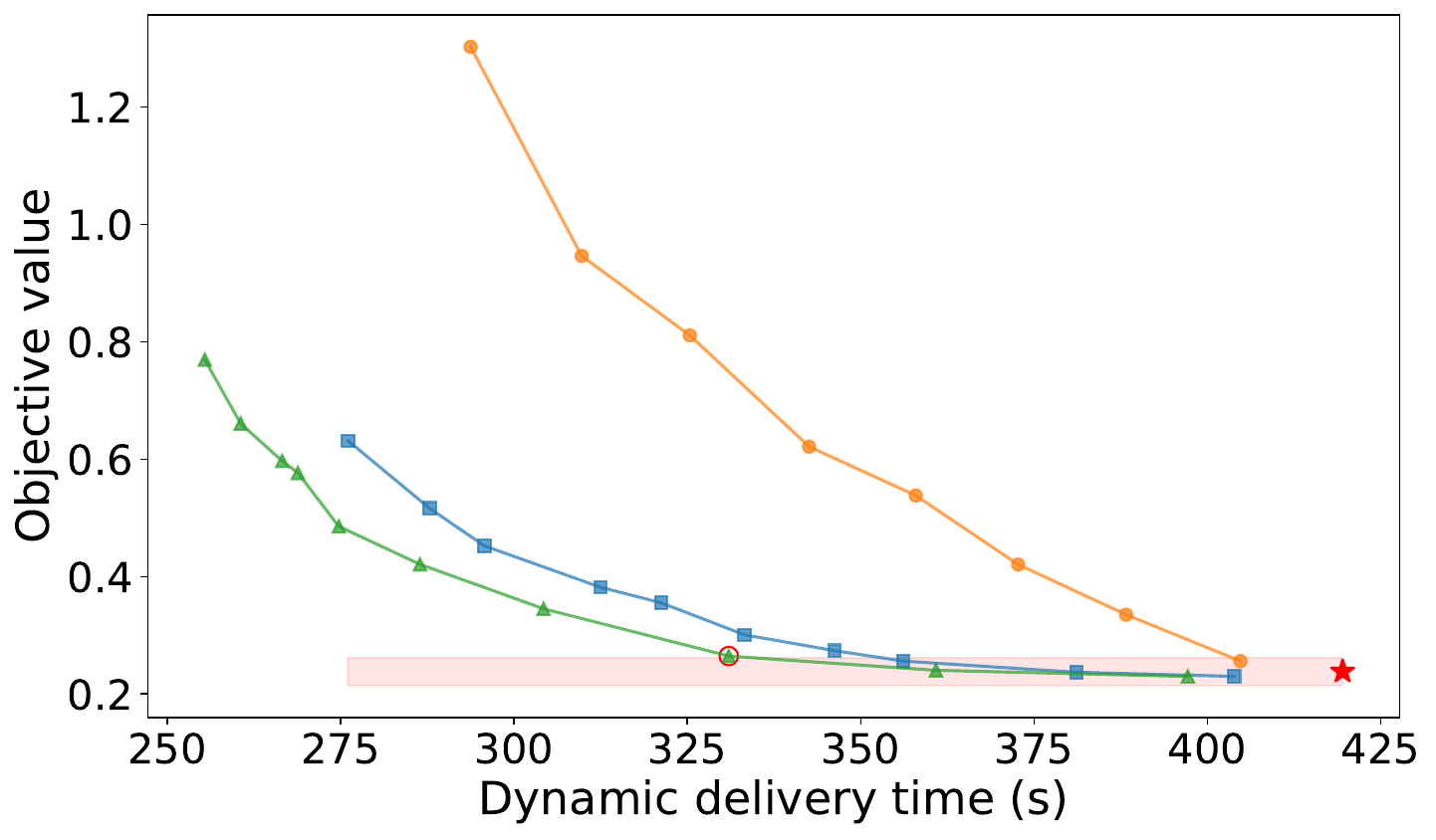} }}%
    \qquad
    \subfloat[\centering OPC 3]{{\includegraphics[width=8cm]{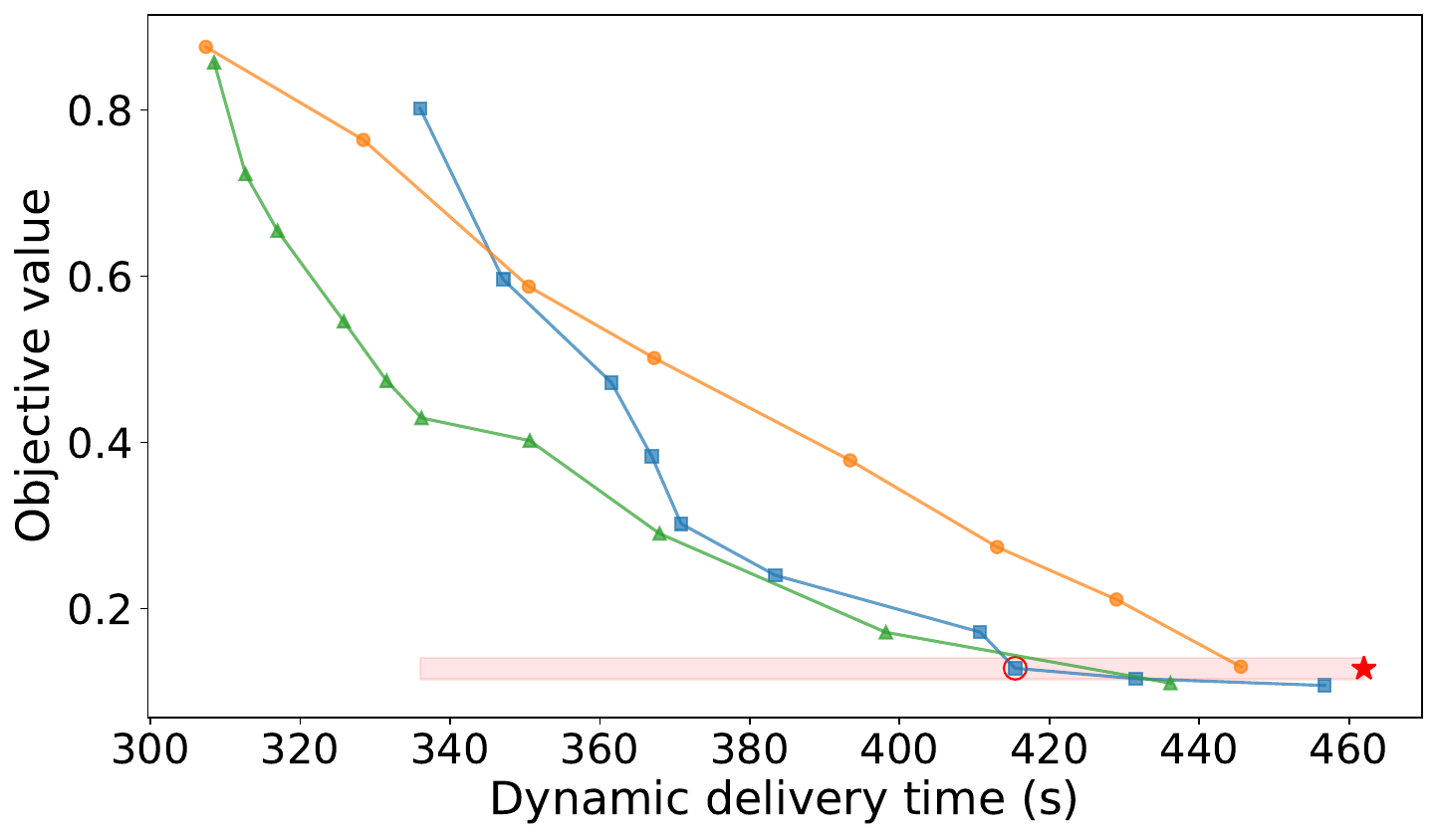} }}%
    \subfloat[\centering OPC 4]{{\includegraphics[width=8cm]{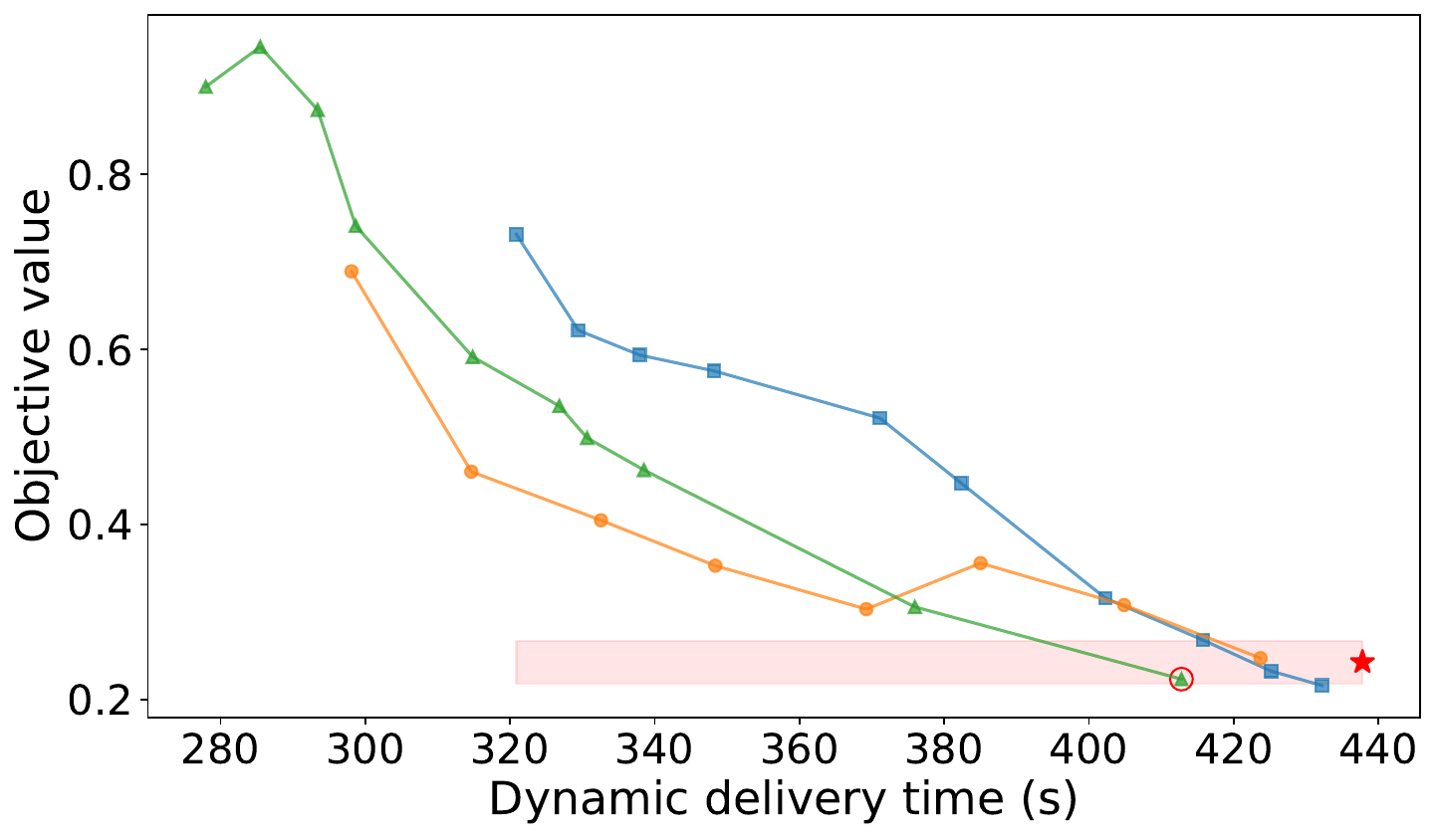} }}%
    \caption{Pareto fronts balancing the objective function value versus \textbf{dynamic beam delivery time} for three energy layers filtering methods for the four oropharyngeal patients. The unrestricted method filters the lowest weighted EL while the SU gap filtering removes in priority the EL around a SU. The SU filtering removes the lowest weighted group of ELs that includes a SU. The red star represents the baseline plan from which we filter ELs. The tolerance band encompasses data points corresponding to plans that deteriorates by maximum 10\% the objective value from the baseline plan. Best compromise is circled in red showing the shortest delivery time with reasonable objective value.}%
    \label{fig6:Pareto_bdt_HAN}%
\end{figure}

\begin{figure}[!ht]%
    \centering
    \subfloat[\centering Lung 1]{{\includegraphics[width=8cm]{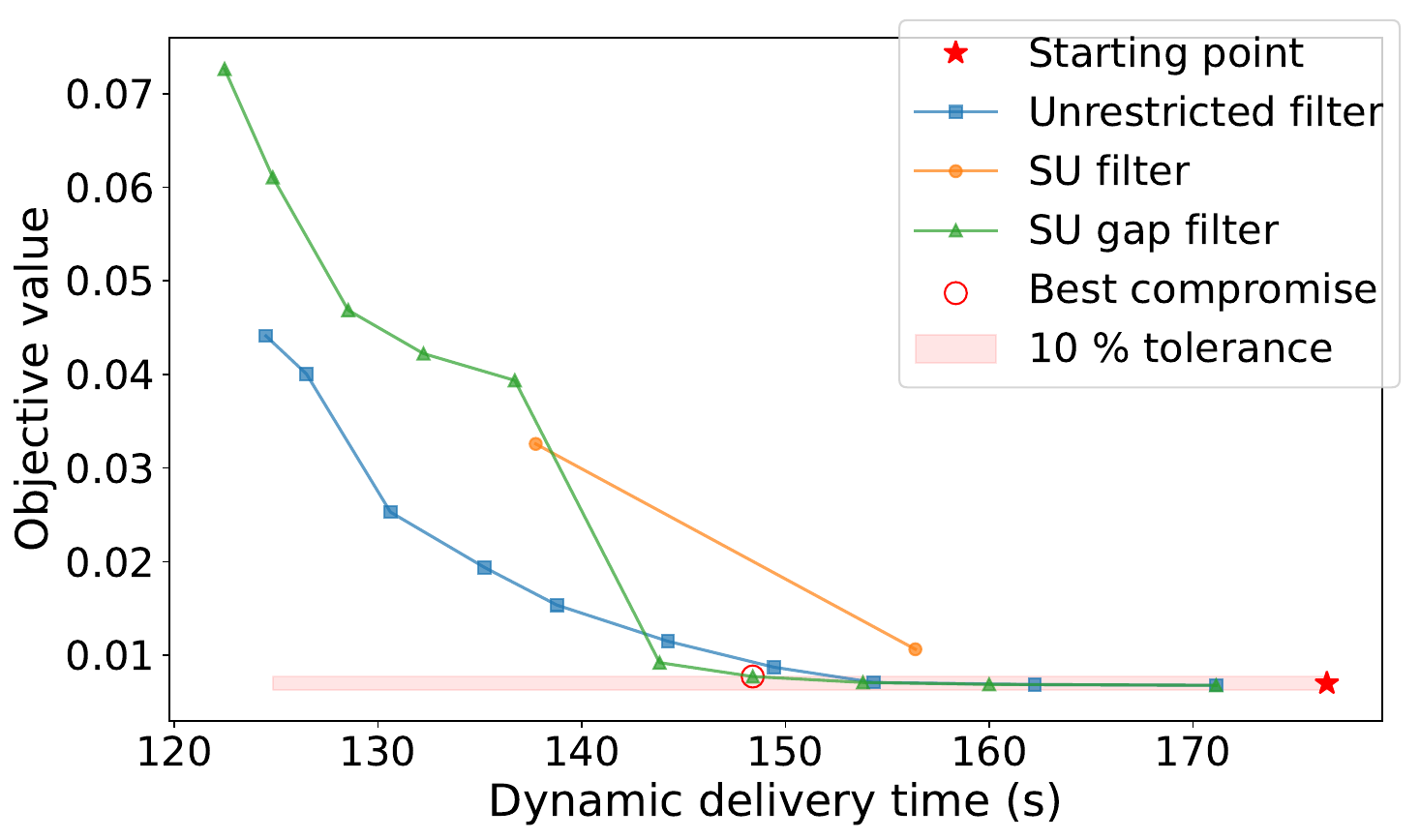} }}%
    \subfloat[\centering Lung 2]{{\includegraphics[width=8cm]{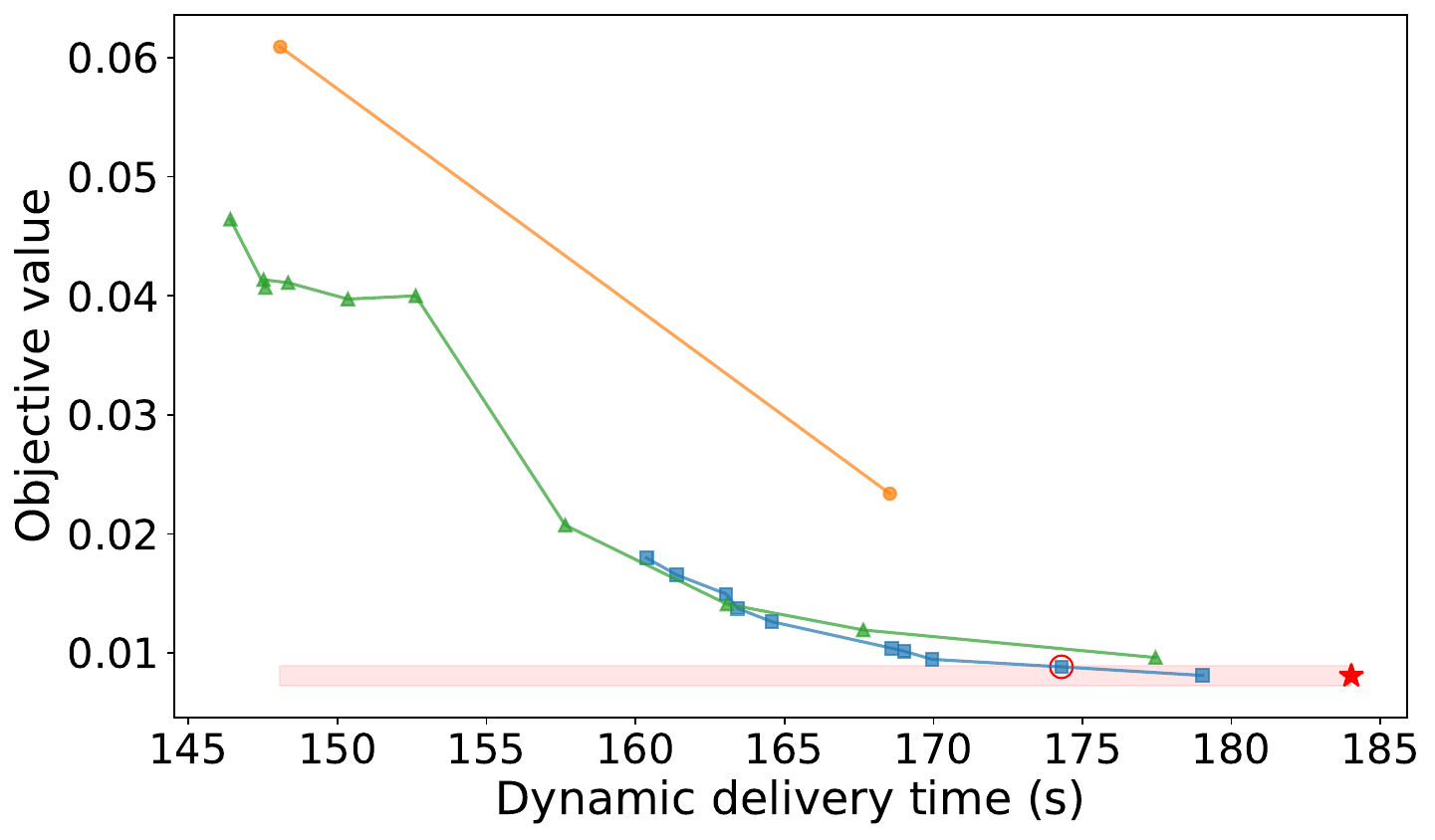} }}%
    \qquad
    \subfloat[\centering Lung 3]{{\includegraphics[width=8cm]{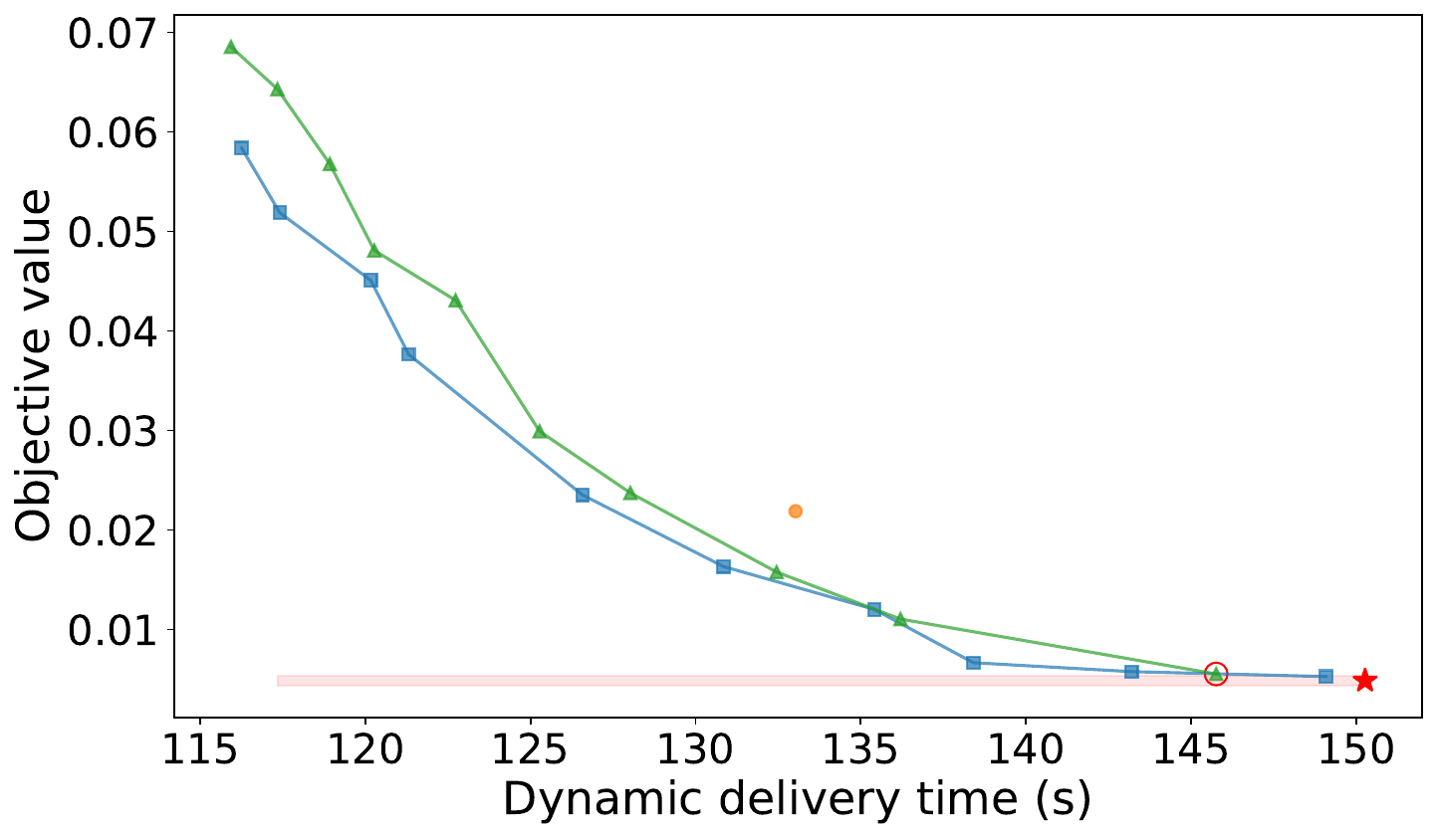} }}%
    \subfloat[\centering Lung 4]{{\includegraphics[width=8cm]{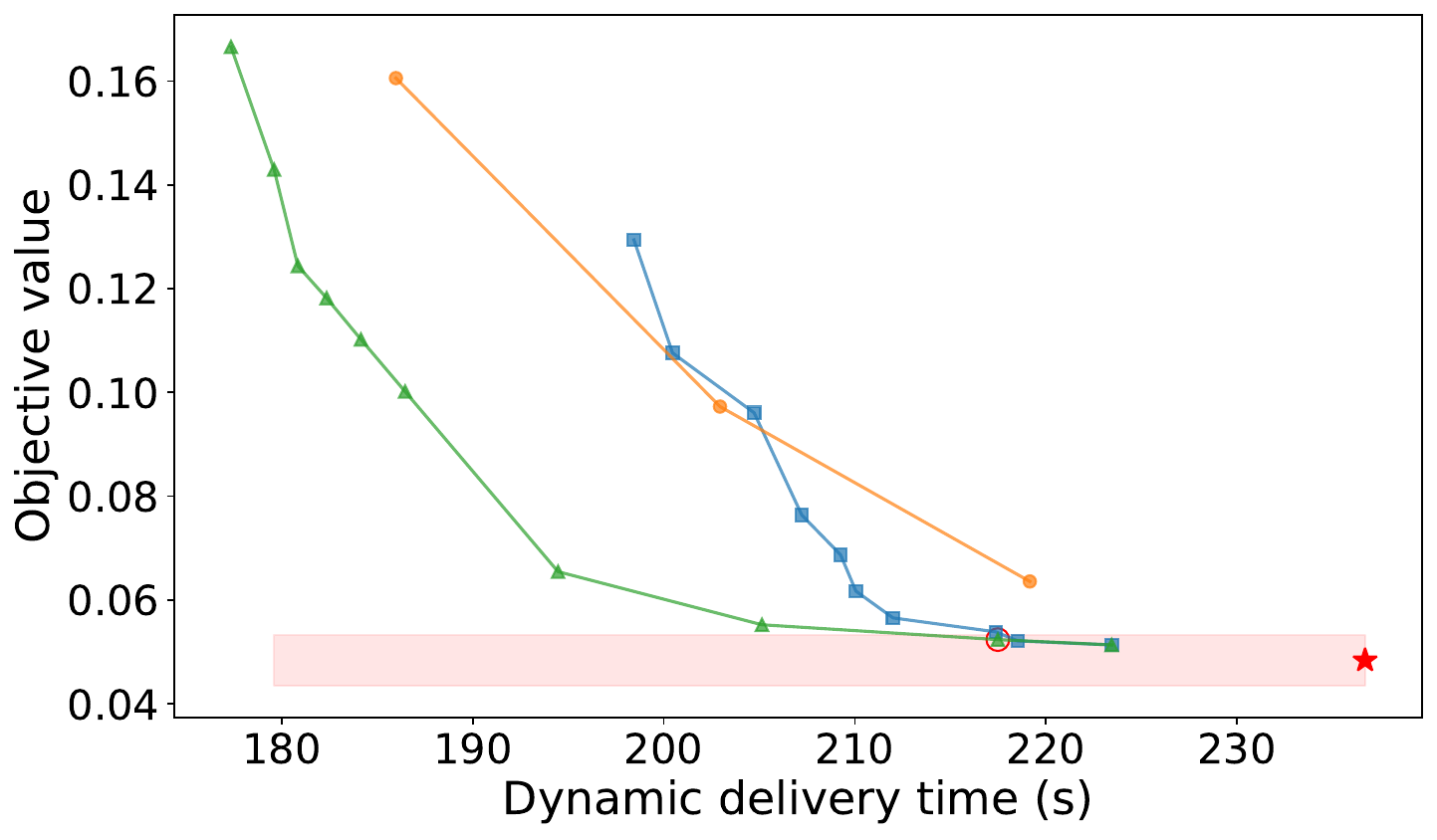} }}%
    \caption{Pareto fronts balancing the objective function value versus \textbf{dynamic beam delivery time} for three energy layers filtering methods for the four lung patients. }%
    \label{fig6:Pareto_bdt_LUNG}%
\end{figure}

The plan quality of the three algorithms is compared for each patient, using the objective function value as a global scalar surrogate of the plan quality. Figures~\ref{fig6:Pareto_bdt_HAN} and \ref{fig6:Pareto_bdt_LUNG} illustrate the evolution of the objective value with the dynamic beam delivery time for the OPC and lung patients, respectively, which are referred to as the true Pareto fronts. The data points are obtained sequentially by running Algorithm S1 as the maximum number of allowed EL or SU in the plan decreases linearly from right to left. Figures \ref{figSM:Pareto_nLayers_opc} and \ref{figSM:Pareto_nLayers_lung} from SM4 provide the related plots with the objective function value as a function of the number of \textit{remaining energy layers} in the plan. The red star represents the optimized plan from which EL or SU are filtered. The filtering process is performed independently for each run, starting from the red star (i.e. the baseline PAT plan). The multiple runs for each algorithm form a front, with a pink band denoting the tolerance range around the initial objective value that should not be exceeded (above). A tolerance value of 10\% is employed in this study. The best compromise that can be found is thus, a solution point with the shortest delivery time achievable within the tolerance band.

Interestingly, the assumption that suppressing more energy layers would lead to a stronger decrease in BDT is proven incorrect. For both cohorts, the SU filtering, despite suppressing more energy layers simultaneously, actually exhibits the smallest reduction in delivery time. On the contrary, the SU gap filtering effectively reduces the beam delivery time with just a small number of suppressed energy layers. For the lung patients, except for lung 1, filtering energy layers entails deteriorating the objective value too much for all proposed methods. However, it can be noticed that the initial number of energy layer in lung patients is much lower compared to OPC patients (see Fig. 1 and 2 of SM4), suggesting a possible correlation with the quality of the filtering results. Referring to Fig. \ref{fig6:Pareto_bdt_LUNG}, it can also be seen that very few runs of the SU strategy were done given the poor results obtained with a single SU filtering.    

The optimal trade-off between plan quality and delivery time can be determined for each patient, by studying the Pareto front within the tolerance band. The SU gap filter proves to be the most efficient for OPC 1 and 2, achieving an effective reduction of approximately 22\% in beam delivery time by suppressing 44 energy layers in the third run, while still maintaining plan quality within the tolerance band. It is closely followed by the simplistic unrestricted filter, which reduces beam delivery time by up to 18\% and 15\% for OPC 1 and 2, respectively. SU gap filtering was also the best compromise found for OPC 4 although it could only decrease the BDT by 6\%. Conversely, the SU filter demonstrates limited improvement, as it could only suppress 1, 2, 1 and 1 SU for OPC 1, 2, 3, and 4, respectively, without exceeding the tolerance zone for plan quality. Consequently, the SU filter shows the smallest gain in beam delivery time. OPC 3 presented a bigger overal CTV, compared to the three other cases and raised tougher planning challenges. The filtering for this patient exhibited better results with the unrestricted filtering method. It achieved a 10\% reduction in beam delivery time in the third run. This smaller reduction in beam delivery time, as the one observed for OPC 4, compared to the two first patients may indicate that more energy layers are required to meet the clinical goals effectively for these particular patients. Nevertheless, SU gap Pareto front in OPC 4 looks like we could have had filtered a few more energy layers while staying in the tolerance band. For lung patients (Fig.~\ref{fig6:Pareto_bdt_LUNG}), unrestricted and SU gap filtering provide very similar results with the Pareto curves being interlaced for the first filtering rounds. The largest BDT reduction relative to the non-filtered plans, while remaining in the tolerance band, are 15\%, 5\%, 3\%, and 8\% respectively for lung 1, 2, 3, and 4. These low numbers result from the very small number of energy layers being filtered, 15 of 141, 7 of 141, 2 of 91, and 13 of 181, respectively. The final best compromises found for each patient are circled in red in each plot.

Figure~4 provides an illustration of the energy layers retained in OPC patients after determining the best trade-offs for each filtering method, compared to the PAT baseline plan. Additionally, relevant metrics are presented to assess the timing results. In particular, the SU filtering demonstrates a consistent pattern of energy layers suppression around 270° for all four OPC patients. It reduced the number of SU from 16 to 15 for OPC 2, 3, and 4, while it filters one more SU for OPC 1 (16 $\rightarrow$ 14). The removal of the second SU for OPC 1 appears to be symmetrically suppressed relative to the first SU. One possible explanation for this behavior is that the optimizer tends to avoid shooting through the shoulders, which may require higher energy levels and deliver higher dose to the body. Another explanation could be related to the position of the primary tumor and also the relative position of the OAR to the tumor. The primary tumor is centrally located for the first three patients while it slightly shifted to the right for OPC 4. In contrast, the unrestricted filter sporadically suppresses energy layers, resulting in holes across each sector. Moreover, this filter tends to remove high energy layers. On the other hand, the SU gap filtering displays a cleaner pattern by sequentially removing energy layers around the SU points. This method generates gaps primarily in between sector transitions. 

\begin{figure}[!ht]%
    \centering
    \includegraphics[width=0.9\textwidth]{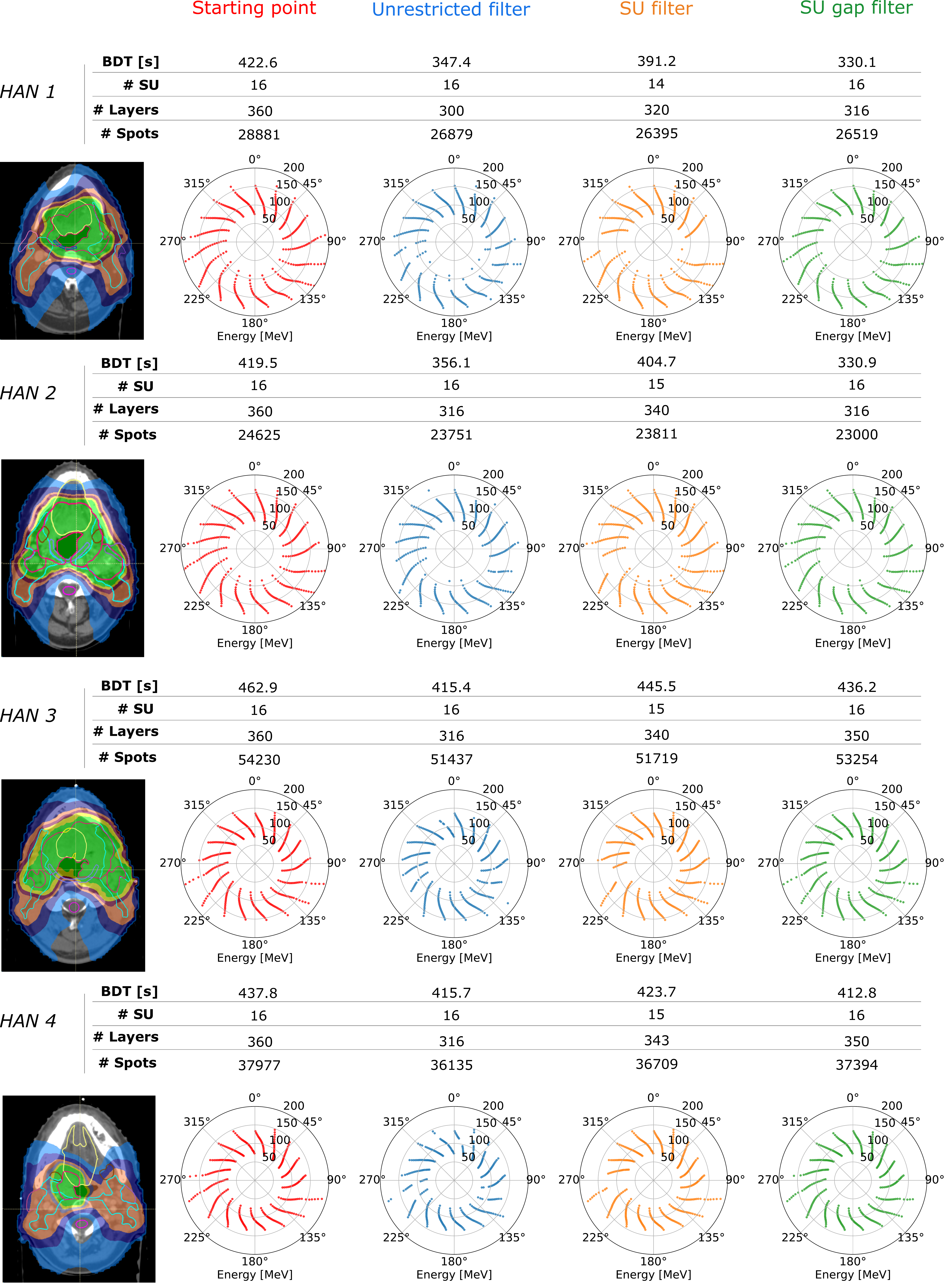}
    \caption{Comparison of energy layers kept in the head and neck plans after best filtering compromise found by the different strategies. Rows display each patient whereas columns showcase the filtering algorithm.}%
    \label{fig6:PolarPlotHAN}%
\end{figure}

\begin{figure}[!ht]%
    \centering
    \includegraphics[width=0.9\textwidth]{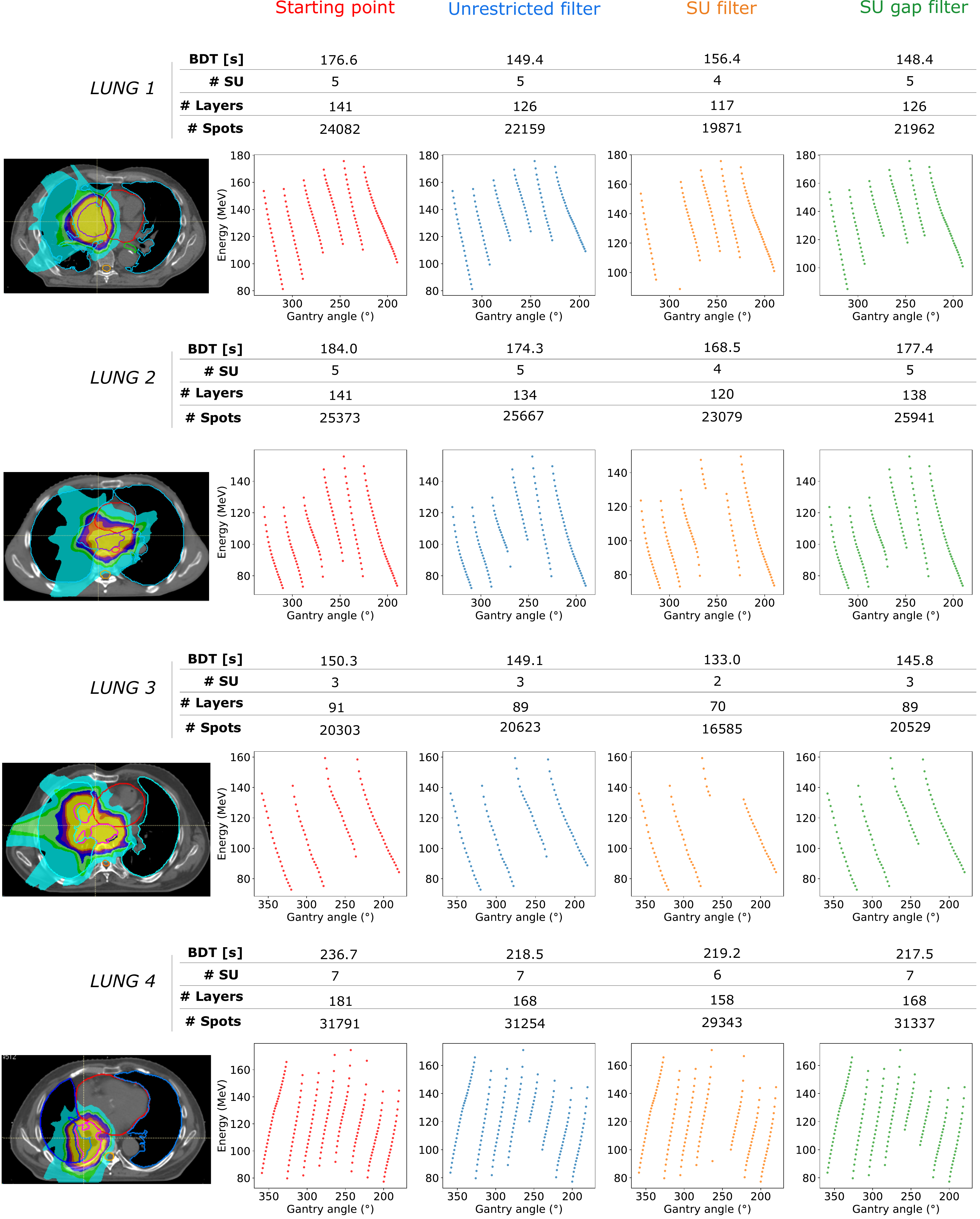}
    \caption{Comparison of energy layers kept in the lung plans after best filtering compromise found by the different strategies. Rows display each patient whereas columns showcase the filtering algorithms.}%
    \label{fig6:ESPlotLUNG}%
\end{figure}

The same analysis was conducted for the lung patients in Fig.~\ref{fig6:ESPlotLUNG}. Given that a relatively small partial arc was used for each plan design (see Table S2), the polar plot representation was replaced with a simple x-y plot representing the energy layers against the angle sequence. As the location and shape of the lung tumor differ for each patient, no obvious pattern can be observed in terms of energy layer filtering. It is also more difficult to see where the energy layers were actually removed as the best compromises only remove a very few of them (except for the SU filtering). 

\begin{figure}
\centering
\begin{subfigure}{.5\textwidth}
  \centering
  \includegraphics[width=.85\linewidth]{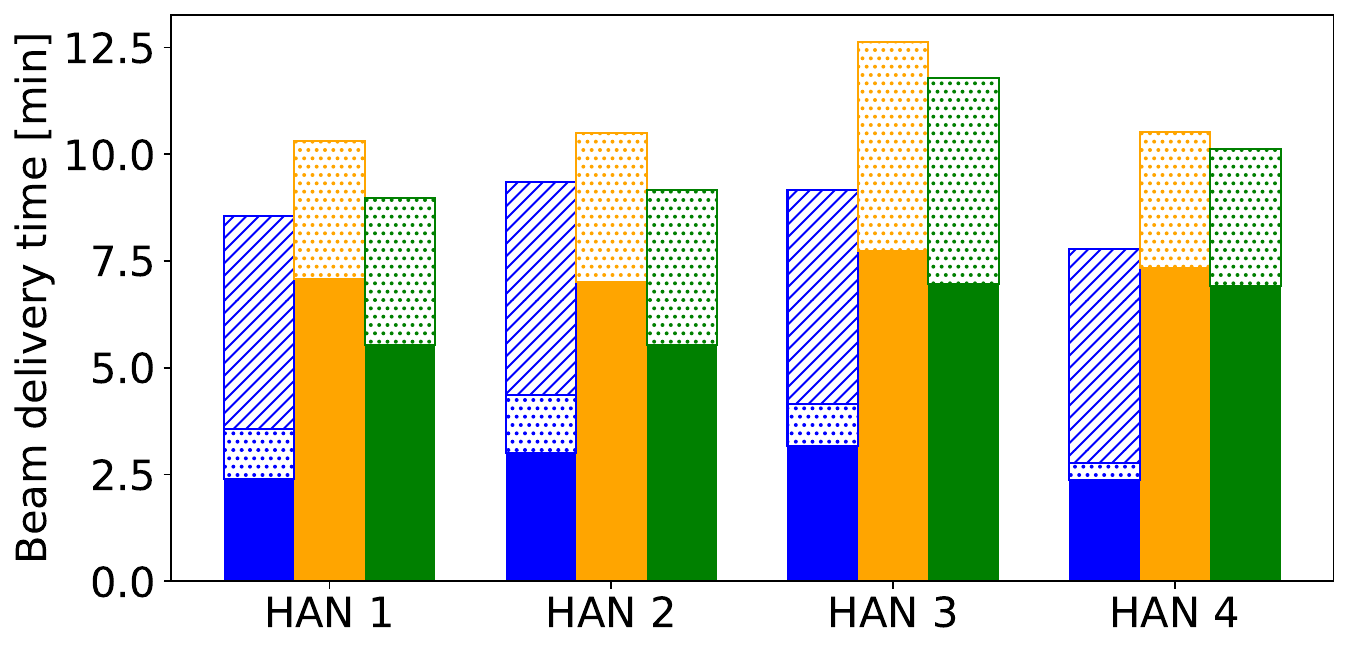}
  \caption{OPC patients}
  \label{fig:bdtHAN}
\end{subfigure}%
\begin{subfigure}{.5\textwidth}
  \centering
  \includegraphics[width=1.23\linewidth]{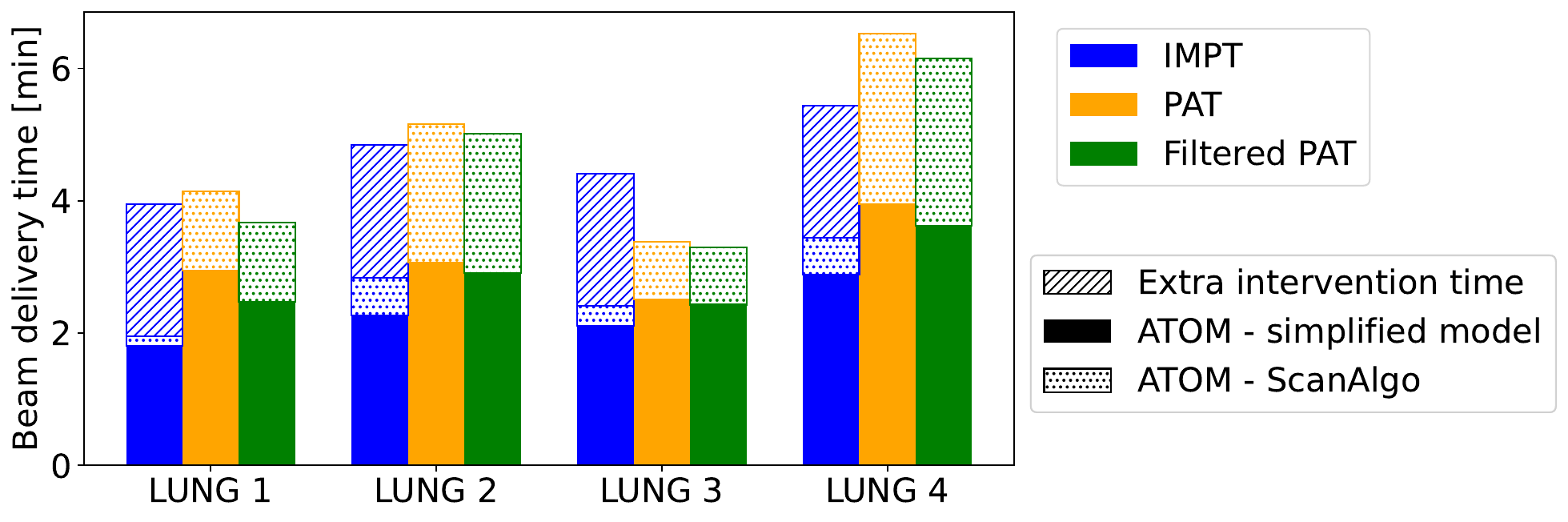}
  \caption{Lung patients} 
  \label{fig:bdtLUNG}
\end{subfigure}
\caption{Comparison of beam delivery times between IMPT, PAT, and filtered PAT plans for the best compromises found for each patient. Filtered PAT employed the SU gap filtering algorithm for patients 1, 2, and 4 whereas unrestricted filtering was applied for patient 3. Plain bars represent a simplified approximation of the static BDT (energy layer irradiation time plus switching time) used in ATOM to which is added (but only for arc plans) the dynamic BDT component as computed by the ATOM simulator for in-between sector transition times. Dotted bars represent the static BDT as obtained for an IBA PPlus system with ScanAlgo combined with (but only for arc plans) ATOM dynamic BDT.}
\label{fig6:bdt}
\end{figure}

After optimizing filtered proton arc plans (filtered PAT plans) for each patient and keeping the best compromise found, the next question that arises is whether these plans can be delivered more rapidly than IMPT plans. In the case of the IMPT plans, the delivery time is computed for each beam individually, resulting in what is refered to as static delivery time, as opposed to the dynamic delivery time of proton arc plans. To enable a fair comparison between static and dynamic delivery timings, one should consider the manual interventions required between each beam delivery in the control room and/or treatment room. This includes beam loading in the console, rotation of the gantry and/or the couch with potential re-imaging step. Moreover, given that range shifters are used for each IMPT beam for the OPC patients, extra caution is needed to avoid any collision. After consulting with two separate proton centers, a simulated delay of 60 seconds is therefore introduced between each beam plus an additional time of 120 seconds required for the couch-kick for each of the IMPT plans. Figure \ref{fig6:bdt} illustrates the delivery time comparison. Note that the comparison indicating faster delivery times for all arc plans than for IMPT plans with the simple ATOM model is valid only when factoring in the (estimated) additional intervention time associated with IMPT plans. For both cohorts, the filtered PAT plans lead to considerable improvement compared to the IMPT delivery time (including extra intervention time). 

For OPC patients, the biggest improvement is observed for OPC 1 and 2 where the filtered PAT plans speed up the delivery by 25\% and 31\%, respectively, compared to IMPT. Regarding PAT baseline plans, they were only able to speed up the delivery by 5\% and 9\% for these patients. OPC 3 and 4 show more resistance for the delivery speed-up but filtered PAT plans still show an improvement of 15\% and 6\% compared to IMPT. However, the estimation obtained with a realistic model from an IBA PPlus machine shows the inverse behavior, with the IMPT plans having shorter delivery times compared to the arc plans. Filtering could help reduce the gap with the IMPT BDT but there is still some margin left for further improvement. Using ScanAlgo BDT, filtering the PAT plans achieved smaller gain in terms of delivery time (compared with the ATOM estimates) with 13\%, 13\%, 7\%, and 4\%, respectively, for OPC 1, 2, 3, and 4. Regarding the lung patients, as three beams are used for the IMPT plans, only two gantry angle changes were taken into account in the delivery time calculation, amounting to one extra minute per IMPT plan when we include the intervention time. Using these numbers and the ATOM simplified model, the filtered plans still present an important BDT reduction of 35\%, 32\%, 40\%, and 26\% compared with the IMPT plans for Lung 1, 2, 3, and 4, respectively. When using scanAlgo to compute the irradiation timings, these numbers change substantially but BDT speed-up is still observed for two patients (Lung 1 and 3).

\smallskip

\subsection{Treatment plan evaluation}

A dosimetric evaluation for each of the best compromises found for each patient of the OPC cohort was carried out and is described in this section. The metrics are compared with the PAT baseline plan and the IMPT plan. Further details and lung results are also provided in Tables 5 and 6 of SM4.

\begin{figure}[!ht]
    \centering 
    \begin{minipage}{.45\linewidth}
        \begin{subfigure}[t]{.9\linewidth}
            \includegraphics[width=\textwidth]{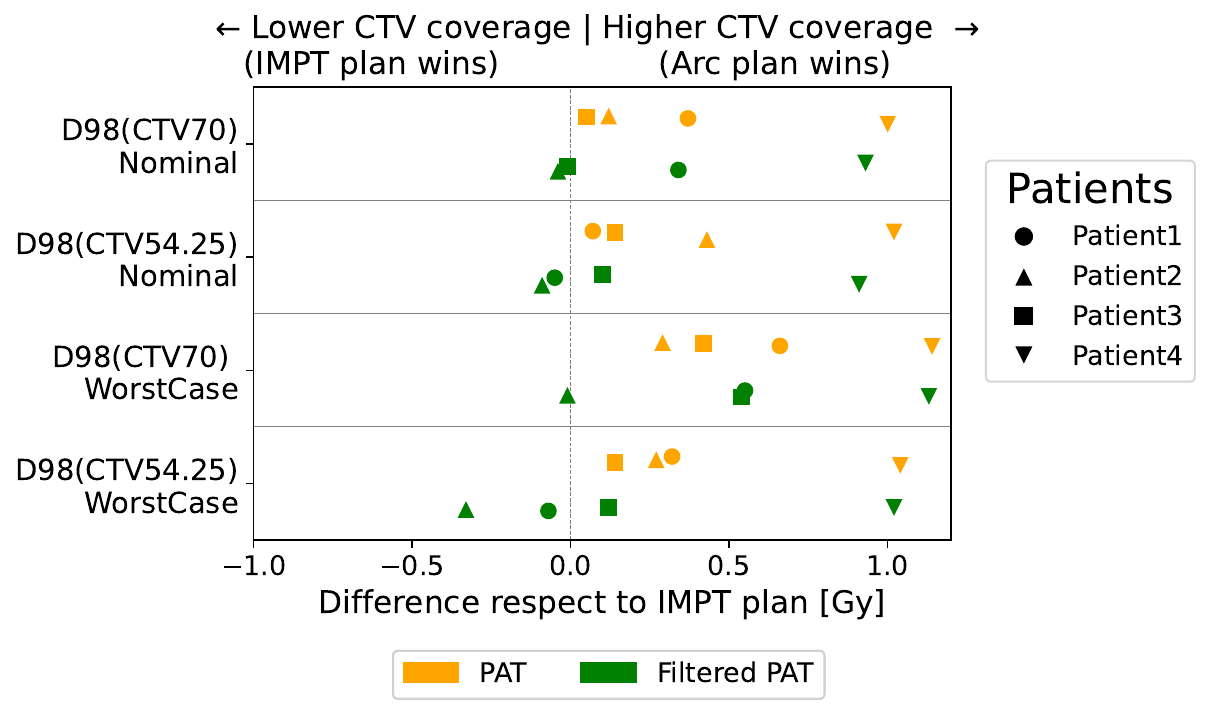}
            \caption{Target coverage}
            \label{fig6:robustEvalCTV}
        \end{subfigure} \\
        \begin{subfigure}[b]{.9\linewidth}
            \includegraphics[width=\textwidth]{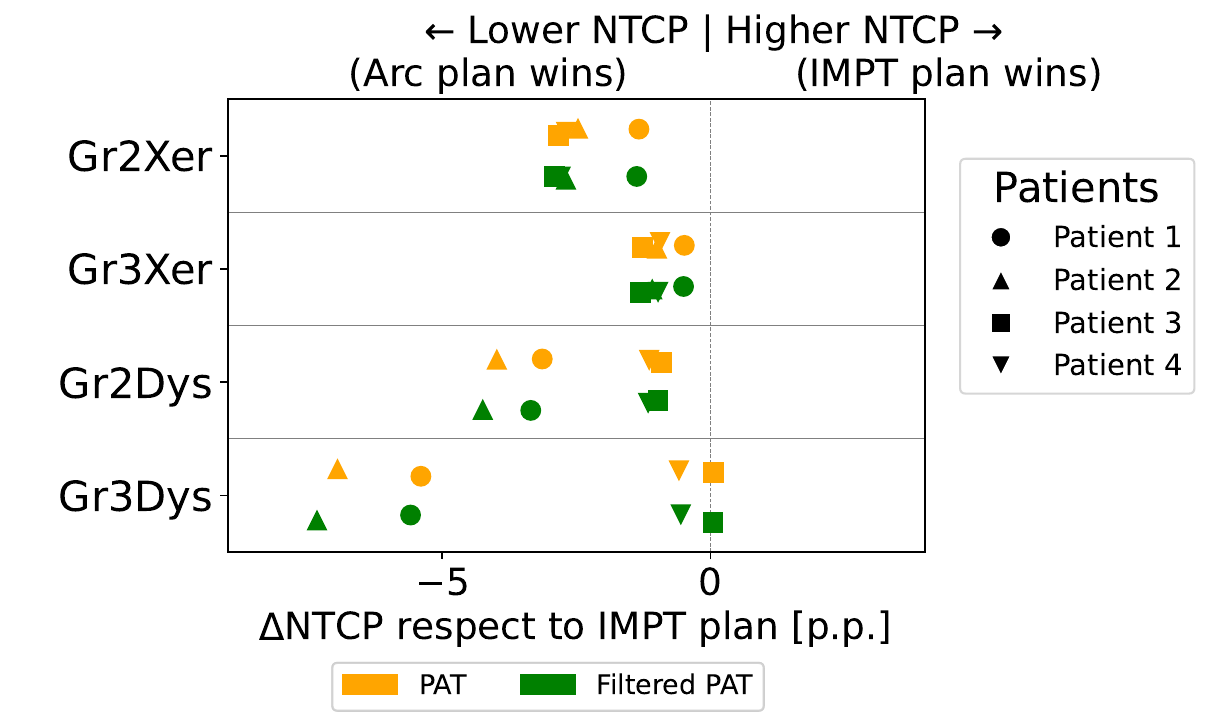}
            \caption{NTCP}
            \label{fig6:NTCP}
        \end{subfigure} 
    \end{minipage}
    \begin{minipage}{.45\linewidth}
            \begin{subfigure}[t]{.9\linewidth}
                \includegraphics[width=\textwidth]{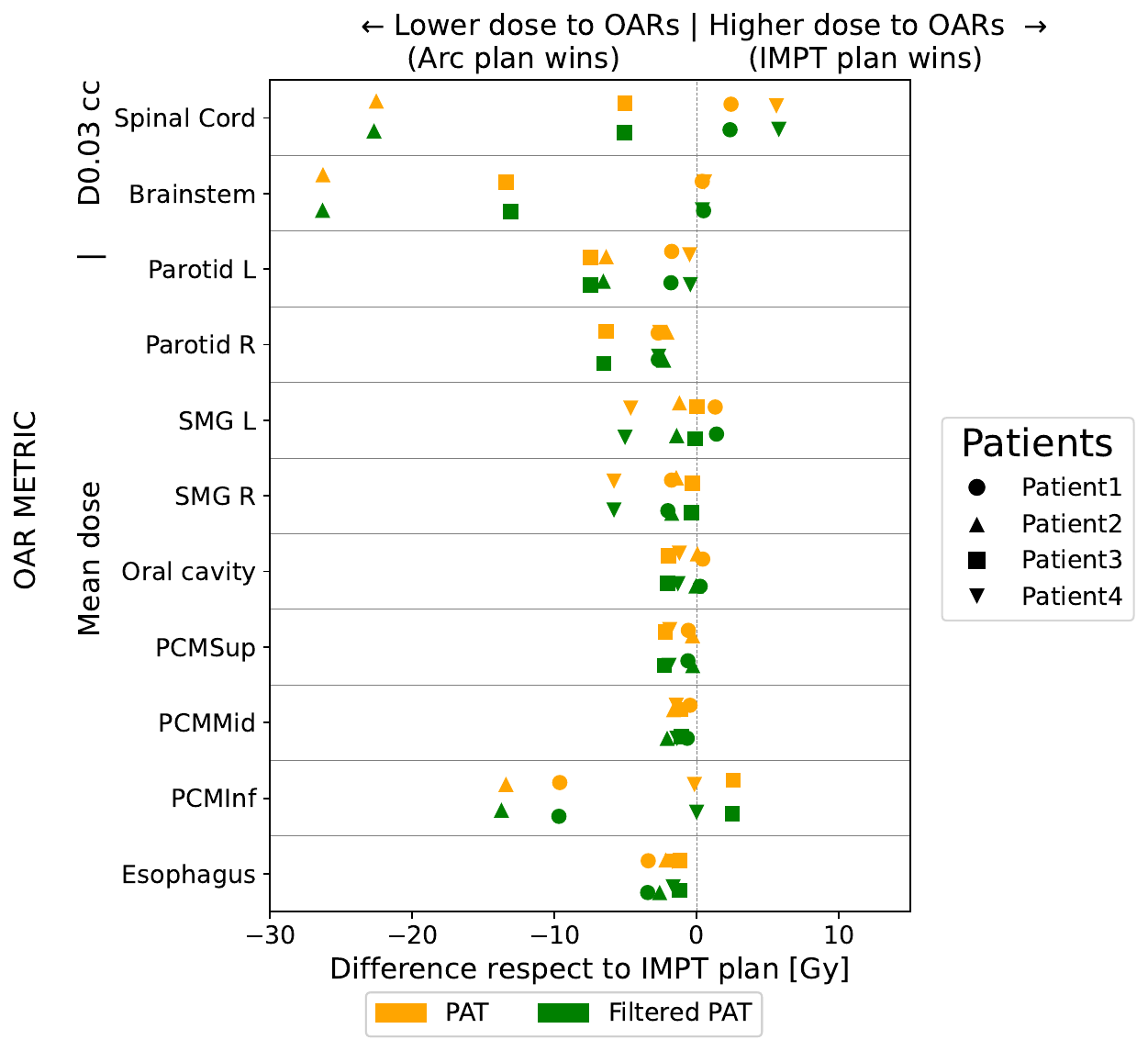}
                \caption{OAR dose}
                \label{fig6:robustEvalOAR}
            \end{subfigure}
        \end{minipage}
    \caption{Differences in (a) target coverage (D98 - dose received by 98\% of the CTV) (b) clinical planning criteria for the OAR (c) NTCP, between arc plans and the IMPT plan. (a) Negative differences imply that the IMPT plan was better than the arc plan in terms of target coverage. (b,c) Negative differences imply that the arc plan was better than the IMPT plan in terms of dose to OAR and probability to develop the side-effect (i.e., less dose to the OAR and lower NTCP). Abbreviations: SMG, submandibular gland; PCM, pharyngeal constrictor muscle. IMPT NTCP values for the ordered four patients: Grade 2 xerostomia = [43.5, 48.7, 53.1, 2.1], grade 3 xerostomia = [12.1, 14.1, 16.1, 11.6], grade 2 dysphagia = [44.2, 43,8, 19.3, 17.2], grade 2 dysphagia = [22.6, 24.4, 3.2, 7.1]. }
\end{figure}

\begin{figure}[!ht]%
    \centering
    \subfloat[\centering OPC 1]{{\includegraphics[width=8cm]{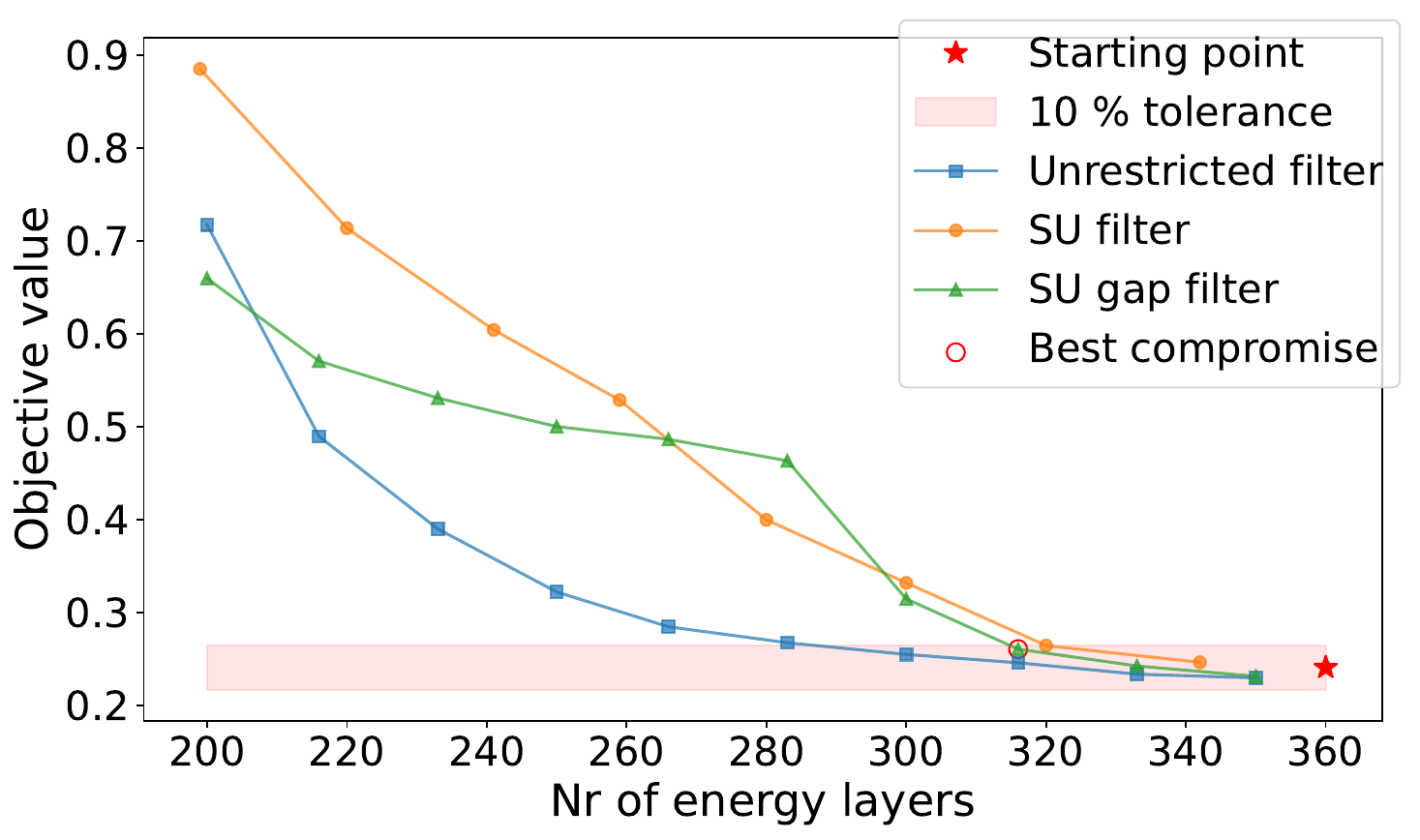} }}%
    \subfloat[\centering OPC 2]{{\includegraphics[width=8cm]{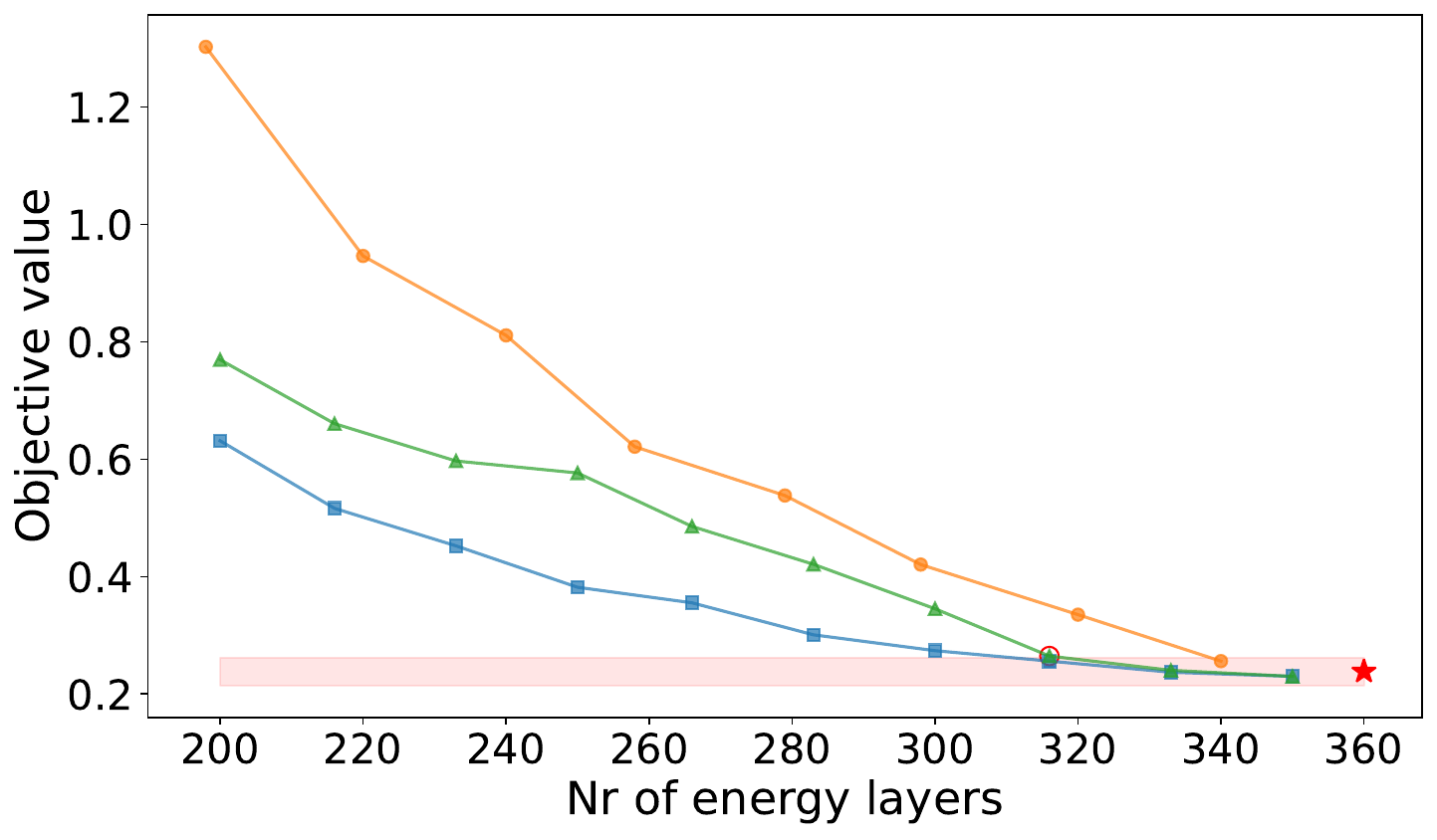} }}%
    \qquad
    \subfloat[\centering OPC 3]{{\includegraphics[width=8cm]{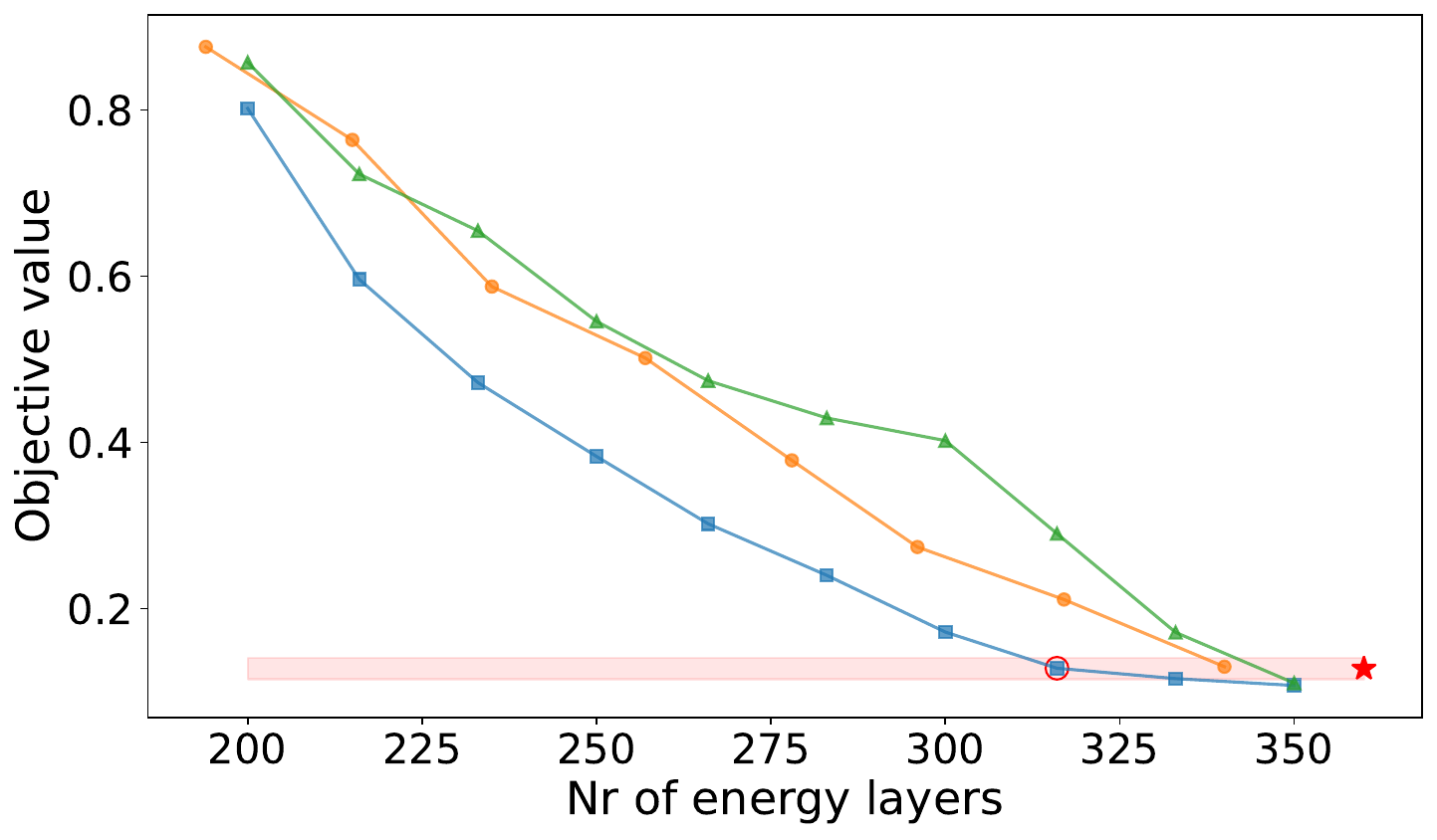} }}%
    \subfloat[\centering OPC 4]{{\includegraphics[width=8cm]{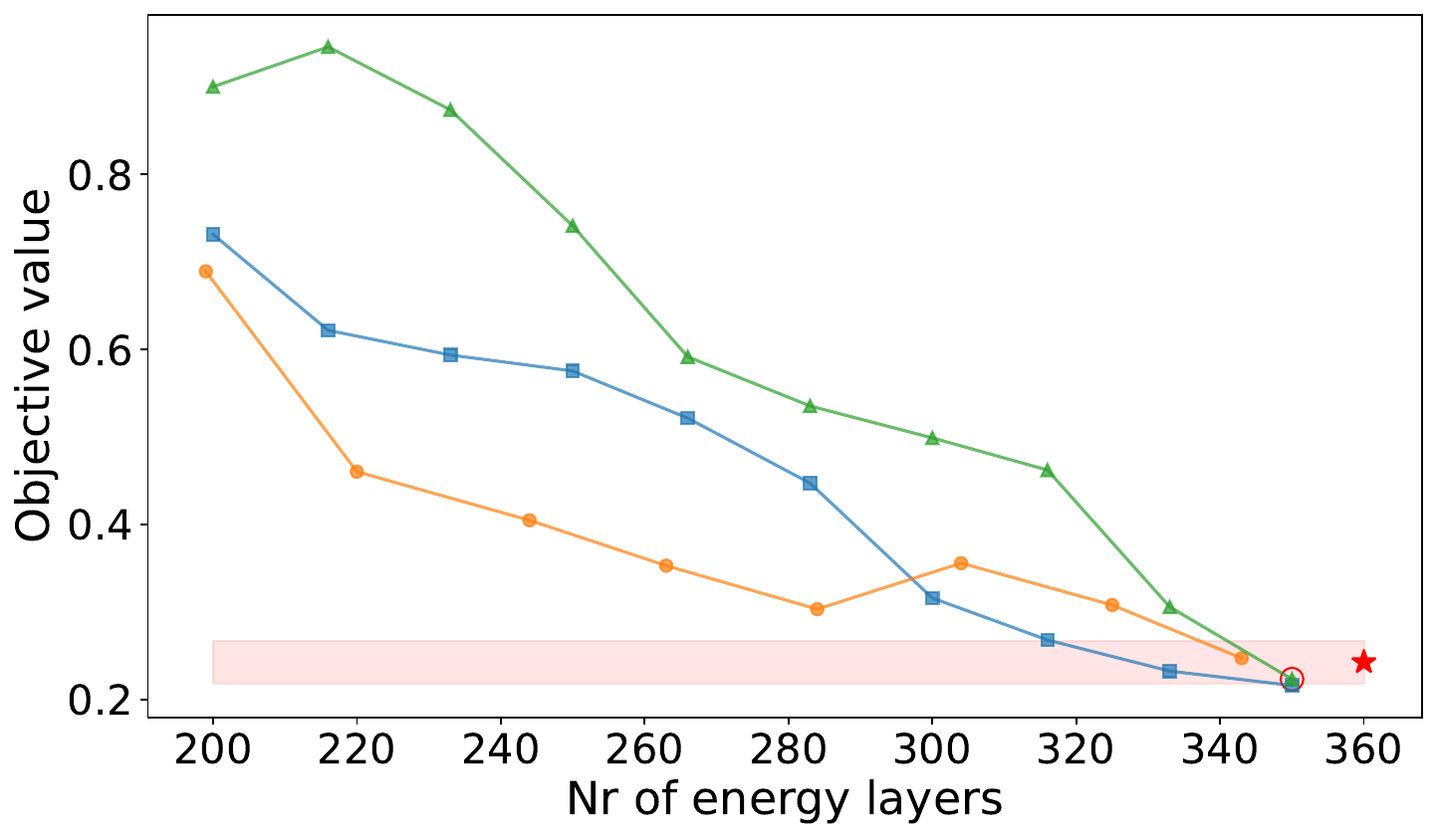} }}%
    \caption{Objective function value versus \textbf{number of energy layers} for three energy layers filtering methods for the four oropharyngeal patients. The unrestricted method filters the lowest weighted EL while the SU gap filtering removes in priority the EL around a SU. The SU filtering removes the lowest weighted group of EL that includes a SU. The red star represents the baseline plan from which we filter EL. The tolerance band encompasses data points corresponding to plans that deteriorates by maximum 10\% the objective value from the baseline plan. Plan quality is deteriorating as more energy layers are filtered from right to left. }%
    \label{figSM:Pareto_nLayers_opc}%
\end{figure}

\begin{figure}[!ht]%
    \centering
    \subfloat[\centering Lung 1]{{\includegraphics[width=8cm]{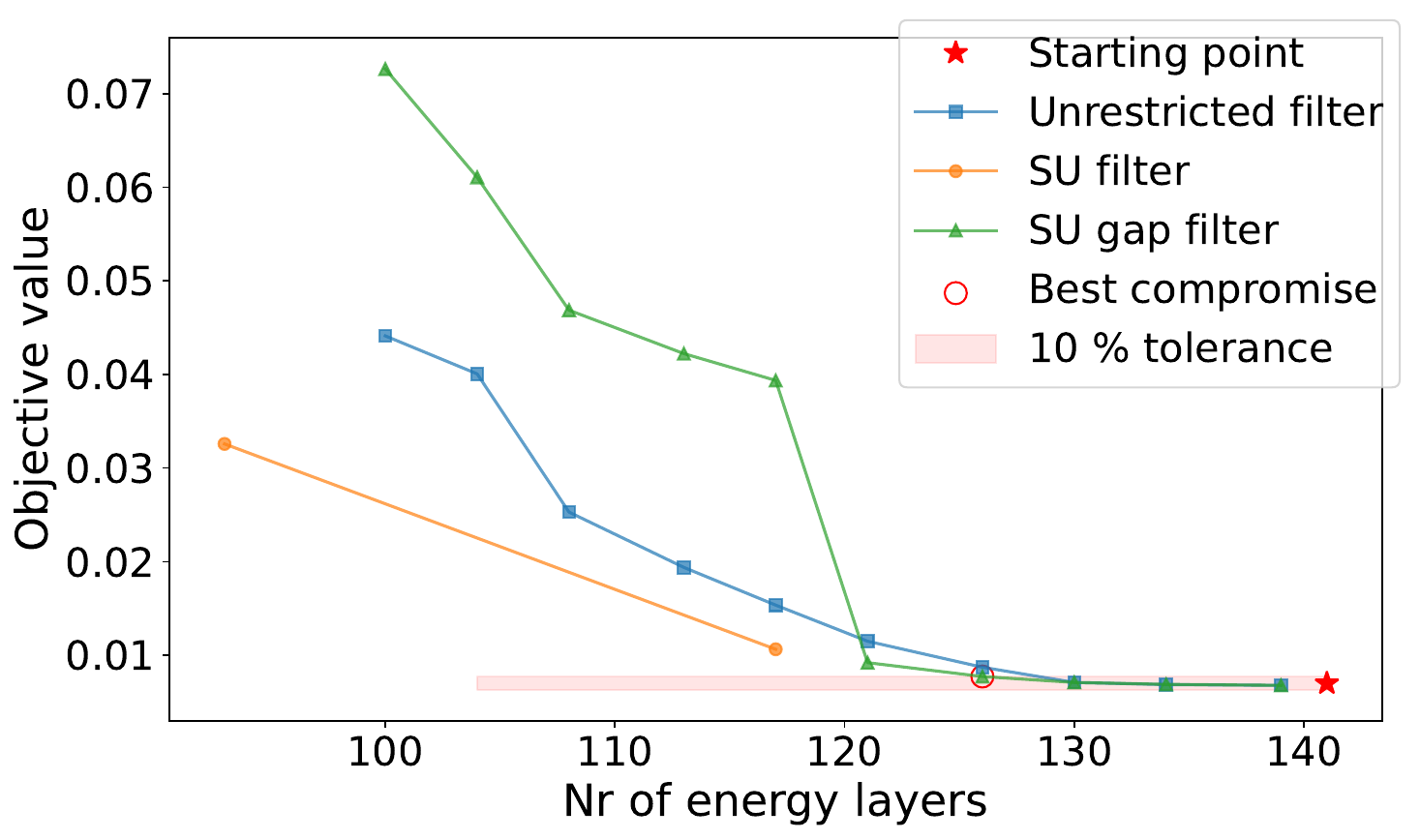} }}%
    \subfloat[\centering Lung 2]{{\includegraphics[width=8cm]{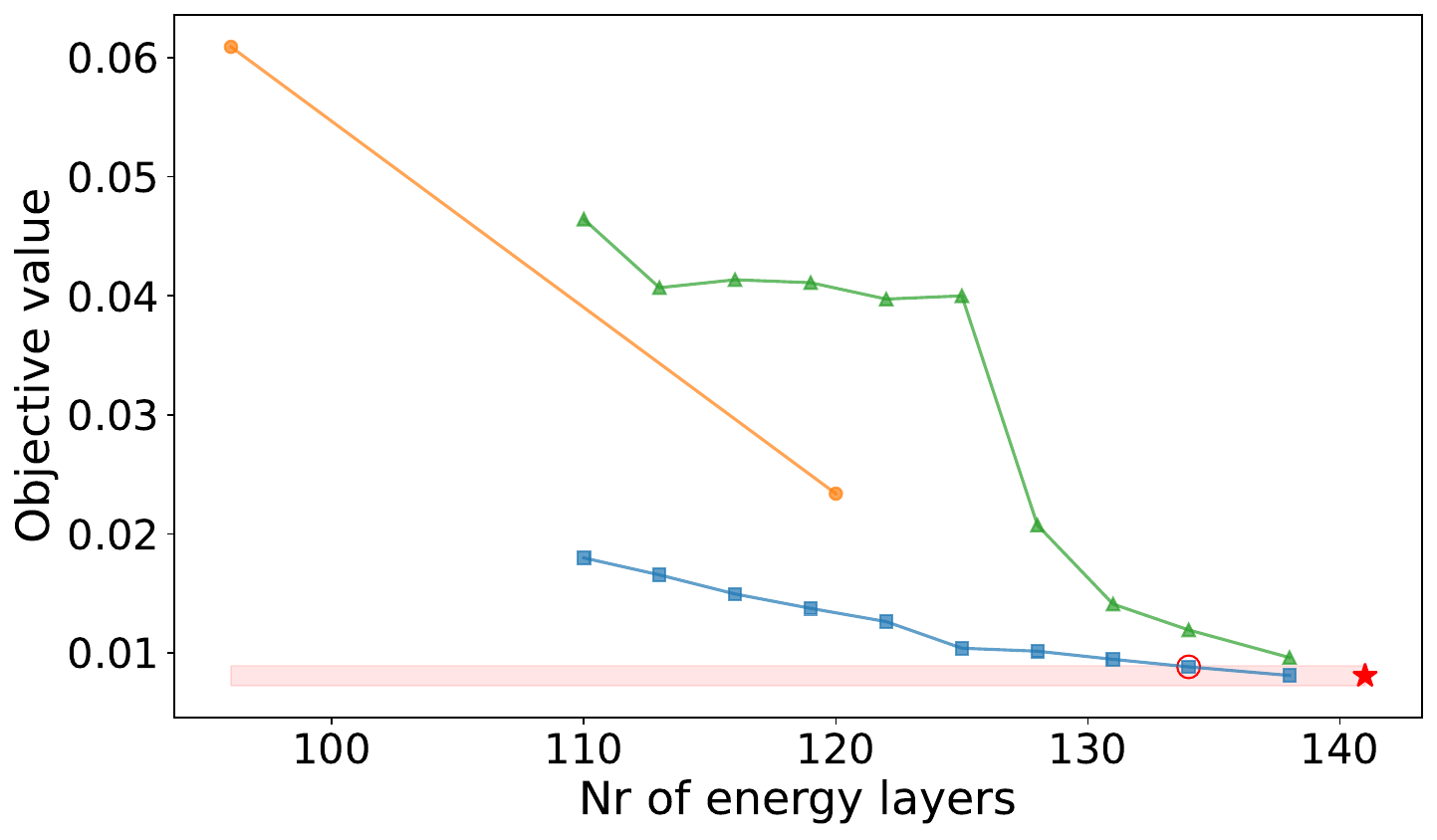} }}%
    \qquad
    \subfloat[\centering Lung 3]{{\includegraphics[width=8cm]{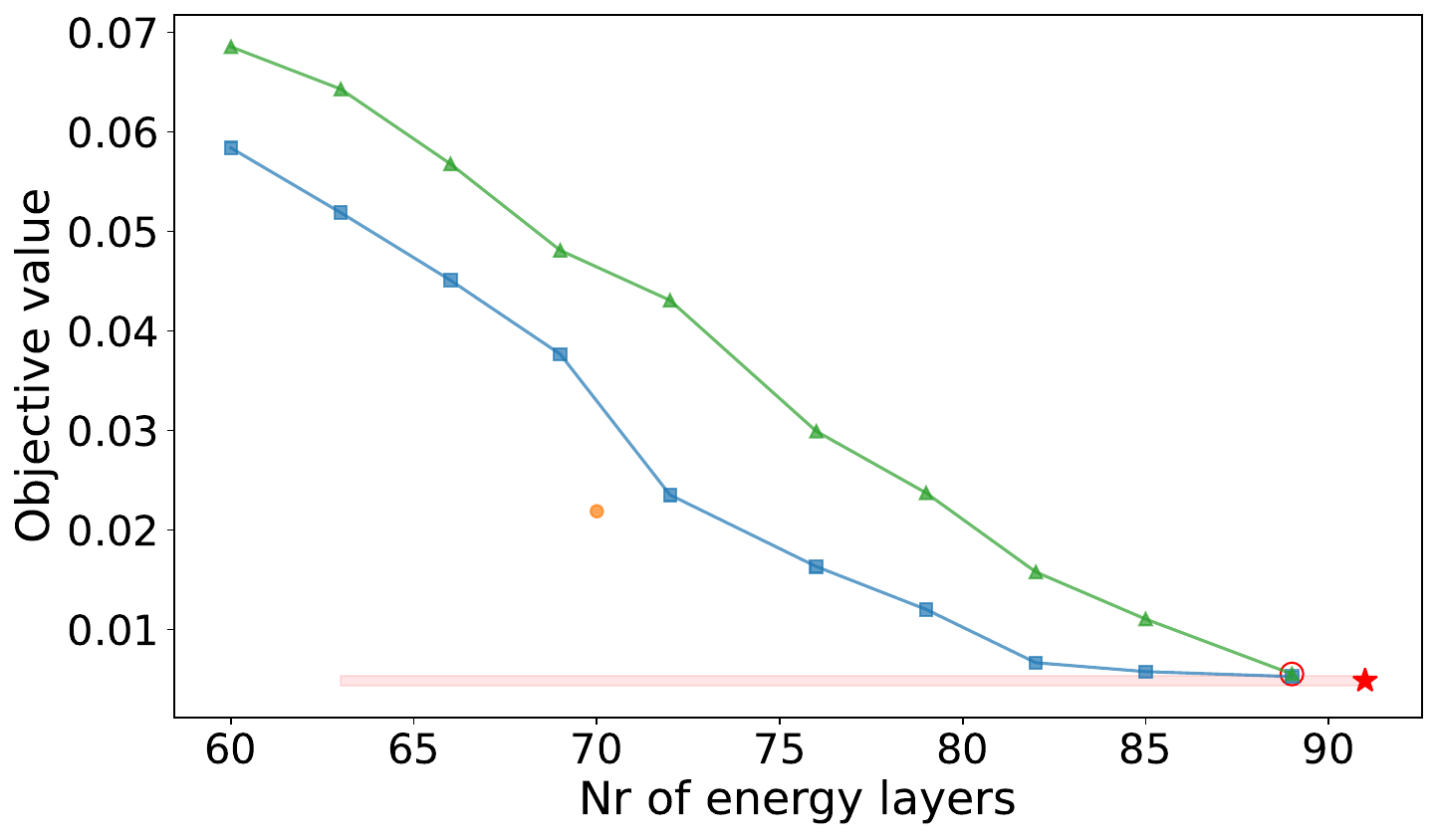} }}%
    \subfloat[\centering Lung 4]{{\includegraphics[width=8cm]{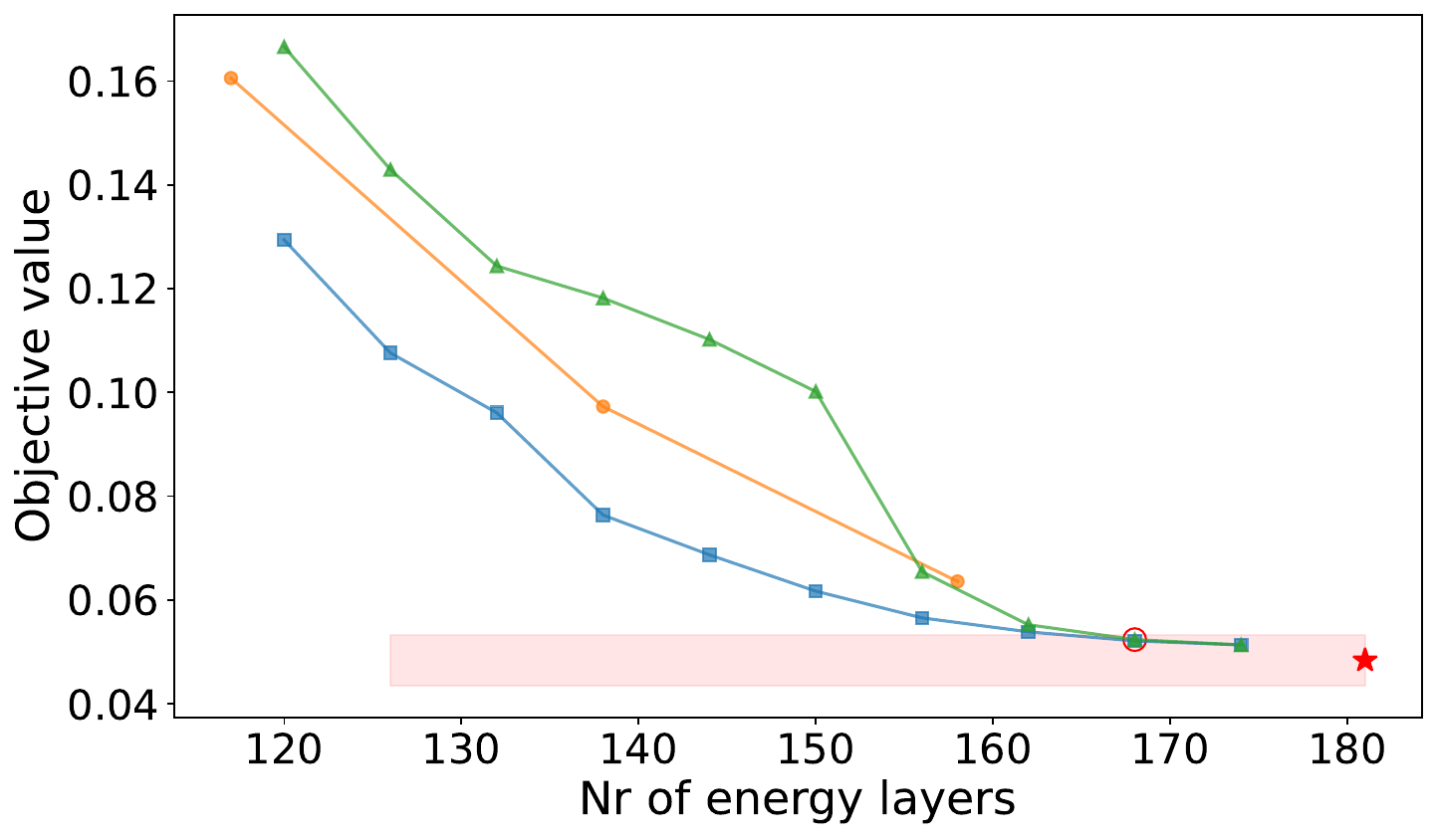} }}%
    \caption{Objective function value versus \textbf{number of energy layers} for three energy layers filtering methods for the four lung patients. The unrestricted method filters the lowest weighted EL while the SU gap filtering removes in priority the EL around a SU. The SU filtering removes the lowest weighted group of EL that includes a SU. The red star represents the baseline plan from which we filter EL. The tolerance band encompasses data points corresponding to plans that deteriorates by maximum 10\% the objective value from the baseline plan. Plan quality is deteriorating as more energy layers are filtered from right to left. }%
    \label{figSM:Pareto_nLayers_lung}%
\end{figure}

Figure \ref{fig6:robustEvalCTV} reports the evaluation results for the CTV targets in terms of D98 in the nominal and worst-case scenario for each patient. The D98 is slightly reduced in the filtered PAT plans compared to the PAT baseline, although the deviation is minimal, not exceeding 1\% for all targets in the nominal and worst-case scenarios (corresponding to at most 0.6 Gy). These small reductions are therefore considered negligible. Both PAT plans yield comparable or better results compared to the IMPT plans with most of the data points displayed on the right-hand side of the plot, bringing higher CTV coverage for arc plans. The clinical goals for target coverage (i.e., D98$\%\geq95$\% Dp), in both high-dose and low-dose CTVs, were met for all the patients in the nominal scenario and the worst-case scenario. 

Figure \ref{fig6:robustEvalOAR} illustrates the DVH metrics differences with respect to the IMPT plan for the OARs included in the clinical goals. Comparing PAT and its filtered version shows a maintained OAR sparing and target coverage, while reducing the number of energy layers. In the transition from IMPT to proton arc modality, only patient 1 and 4 show a slight increase in dose to the spinal cord and brain stem but still under the limit. However, average doses in organs generally decrease (85\% of all data points lie on the left-side of the plot), which is highly favorable for arc modality in OPC patients. Table S5 reports details on the plan quality metrics supporting that filtered plan do not compromise plan robustness. \\

Figure \ref{fig6:NTCP} displays the difference in NTCP relative to the IMPT plan ($\Delta$NTCP = NTCP$_{\text{arc}}$ - NTCP$_{\text{IMPT}}$). The arc modality substantially decreases the NTCP for all reported side effects, including grade 2 and 3 xerostomia and dysphagia. This finding further supports the adoption of PAT for the treatment of this particular cancer site. Moreover, patients 1 and 2, selected for their high NTCP in the IMPT plan, show a reduction of approximately 5 and 7.5 percentage points (p.p.) in NTCP for dysphagia grade 2 and 3, respectively. Similarly, patient 3, chosen for its elevated xerostomia NTCP value, demonstrates a decrease of around 2.5 p.p.~in NTCP for the proton arc plans, further highlighting the advantages of PAT.

\section{Discussion}
In this study, three post-processing energy layer filtering algorithms were presented. They strive to accelerate the delivery of proton arc treatment plans: the unrestricted filtering, the SU filtering, and the SU gap filtering. These algorithms were designed to suppress the low-weighted energy layers or partial sectors in the arc plans generated with ELSA. Among the three methods, the SU gap filter was proved to be highly efficient in reducing the dynamic delivery time by suppressing a minimal number of energy layers. This primarily stems from the inherent design of ELSA plans, built with almost equally spaced sectors and energy layers. Filtering energy layers around a sector transition is intuitively efficient as it gives the system some additional time to make the energy switch with minimal gantry braking. At first glance, this method therefore only works for specific proton arc design, that limits the angular spacing to make an energy layer switch. In fact, this is the case in most proton arc plans since they are generated with a fixed angle spacing. However, the SU gap is not parameter-free as it requires the knowledge of machine-specific parameters (Eq.~\ref{eq6:SUGap}) to know maximum how many energy layers one can remove around each sector transition. The unrestricted filter achieved similar results to the SU gap, although it required more energy layers to be suppressed for the same reduction in BDT in the OPC cohort. This technique is the simplest to set up and was already identified as an efficient means to filter proton arc plans \cite{ding_spot-scanning_2016,battinelli_proton_2019}, but rather focused to have exactly one energy layer per angle to enable efficient dynamic delivery. Regarding the SU filter, despite removing a large number of energy layers (partial sectors), it was the least effective in reducing the BDT and even showed unacceptable results for the lung treatment site (i.e. far outside of acceptable tolerance range). It should therefore probably be discarded from any future implementation. Although these techniques were applied in combination with ELSA for this particular study, they could likely be applied to any proton arc plan, offering particular advantages for plans generated using an energy layer pre-selection method that lacks dose information. This hybrid optimization makes a good compromise between rapid pre-selection algorithms and comprehensive dose-based optimization \cite{gu_novel_2020,zhang_energy_2022,wuyckens_treatment_2022} for PAT treatment planning, as the post-processing filtering incorporates dose information. Overall, the results of the study show that it could be useful to implement both unrestricted and SU gap filtering or even a combination of both strategies in a clinical TPS

Pareto plots, balancing the objective value against the BDT, revealed that the SU gap filtering algorithm often provided a good trade-off between plan quality and delivery time. Further removal of energy layers resulted in a serious increase in the objective value, correlating with a degradation in plan quality that is more sensitive compared to the unrestricted filtering. In the OPC cohort, 2 patients were harder to accelerate. Actually, both patients had much higher initial number of spots, attesting for their more complex tumor geometry. The same can be observed for the lung patients where plans could hardly be filtered without rapidly deteriorating the plan quality. The fact that partial arcs were used to design the lung plans may also have contributed to the difficulty to filter them. A lower number of energy layers was initially available for each of the lung patients. For instance, Lung 3 only had 91 energy layers to start with, as a gantry angle spacing of 2° over half a full arc was sufficient to reach robust target coverage. As a consequence, only 2 energy layers could be filtered without compromising the quality with the SU gap filter, limiting the BDT improvement to only 3\% and gaining 5 seconds by filtering. This observation leads to the conclusion that the filtering techniques may be found to be more useful when the initial number of energy layers is sufficiently large, for instance for complex treatment site geometry with many OARs involved, as seen with OPC patients. Notice that the choice of a 10\% tolerance criterion for the objective values was arbitrary, and further investigation is needed to determine the actual threshold beyond which plan degradation becomes unacceptable and/or robustness is no longer maintained. Moreover, additional iterations may be necessary after the post-processing filter, depending on the number of energy layers removed from the plan. 

While the numbers of energy layers in the IMPT and PAT plans for the lung patients are about the same, the PAT plans for the OPC patients necessitate a much higher number than the IMPT plans. This difference might arise because it is more challenging to combine beamlets from each energy layer in the arc, ensuring an adequately spread-out Bragg Peak to cover the complex targets. It is also important to understand that both the plan quality and BDT heavily rely on the preliminary energy selection. The angular sampling and sector count may not be optimal and could potentially benefit from adjustments that are not purely geometry-driven. Yet, incorporating dose-based objectives into the initial energy layers choice can be complex due to the size of the proton arc problem \cite{wuyckens_treatment_2022}. Hence, dose-based filtering methods paired with fast energy layer selection are a good candidate to address the shortcomings of purely geometry-based energy layer choices.


This study presents BDT results obtained using the ATOM simulator \cite{wase_optimizing_2024}, but it is important to approach these results with caution, as ATOM uses a simplified layer delivery and layer switching model that might not correspond to the exact behavior of most systems. To address this uncertainty, we conducted a benchmark comparison by incorporating BDT data from an IBA PPlus system. While there were some differences in the specific numerical values related to BDT reduction when filtering treatment plans between the two models, we observed that the overall trends remained consistent. The disparity can be attributed to the fact that the IBA ScanAlgo system included more static time in its estimates, a factor directly influenced by the choice of model and parameters. Additionally, the dynamic BDT component, associated with rotation movement, was still simulated using ATOM. This is because, currently, the validation of any dynamic BDT model cannot be provided as we still await for the first dynamic arc product integration in clinical settings. Despite these considerations, we do not anticipate a substantial impact on the overall reliability of the comparative analysis if a more refined dynamic model were to become available. Furthermore, it is worth noting that the current body of literature in PAT planning often employs even simpler approximations of BDT with no consideration for acceleration or jerk using in-house models \cite{liu_novel_2020}. In this context, ATOM may represent one of the most advanced algorithms for computing dynamic BDT and, most importantly, it is available as open-source. In any case, BDT will inevitably vary for each system in relation to the delivery assumptions, and yet most of the conclusions of this study should remain valid.

Another key assumption made for the BDT estimation relates to the IMPT plan. Extra time between each static beam delivery was factored to ensure a more fair comparison with the arc plans. After consulting with two separate proton centers about this additional time, the lowest values were chosen for this study (i.e. faster IMPT delivery). This suggests that the difference in BDT between IMPT and arc plans could be even more pronounced. Moreover, the filtered arc plans further reduced the BDT, attaining up to a 30\% (OPC) and 40\% (Lung) speed-up in comparison to IMPT (using ATOM estimates), emphasizing the utility of the proposed method. Nevertheless, IBA scanAlgo BDT estimates are more pessimistic, hence the proton arc plans usually take longer to be delivered compared to ATOM estimates. While filtering narrowed the gap towards IMPT BDT, there is still room for further optimization in reducing the delivery time of proton arc plans.

The plan evaluation and robustness analysis demonstrated that the filtered PAT plans maintained target coverage with negligible deviations from the PAT baseline plans in both nominal and worst-case scenarios for both patient cohorts. Regarding the OPC results, comparisons with the clinical IMPT plans, generated by an experienced dosimetrist, showed that both the non-filtered and filtered PAT plans outperformed the IMPT plans in most clinical goals, as observed across all four OPC patients. Additionally, the study of NTCP for dysphagia and xerostomia further supported the superiority of PAT for oropharyngeal cancer. These findings concur with existing publications on PAT in this disease site \cite{liu_improve_2020,de_jong_proton_2023}. On the other hand, the arc benefit is less obvious for the lung cohort (see Table S6) although the OAR dose metrics do not suffer from the filtering process.

Nevertheless, despite the complexity of proton arc plans, the implementation of a post-processing algorithm that filters out the lowest MU-weighted energy layers can still have a positive impact on treatment delivery. Even a small reduction in BDT could potentially enhance patient throughput in clinical settings. Furthermore, the integration of delivery-accelerating strategies could be seamlessly implemented into existing treatment planning systems. It is worth noting that post-processing filtration, including the subsequent short re-optimization, only takes around 10 and 15 minutes for OPC and lung patients evaluated in this study (with 5 and 12 minutes for OPC and lung, respectively, which were dedicated to pre-calculating the dose that in principle could be cached). This relatively short running time might fall within an acceptable range of delay for proton treatment planning.

Finally, it is important to acknowledge the limitations of this study, including the small sample size of each cohort and the absence of statistical analysis. Nevertheless, this study provides a proof of concept and lays the groundwork for future investigations.

\section{Conclusion}
In conclusion, this study introduces energy layer filtering as a post-processing step of the dose optimization to accelerate the delivery of proton arc treatment plans. The results demonstrate that these filters effectively reduce beam delivery times while maintaining plan quality, especially when the initial number of energy layers is high. These methods hold particular promise when combined with fast energy layer pre-selection techniques for designing arc plans that lack dose information for their selection. The findings support the use of PAT in reducing organ-at-risk toxicity and improving local tumor control in OPC cancers. Further research is needed to optimize proton arc planning and understand the underlying requirements for energy layer selection. The proposed filters have the potential to enhance clinic efficiency and patient throughput without compromising treatment quality.

\section*{Acknowledgments}
Sophie Wuyckens is funded by the Walloon Region as part of the Arc Proton Therapy convention (Pôles Mecatech et Biowin). Computational resources have been provided by the supercomputing facilities of the Université catholique de Louvain (CISM/UCL) and the Consortium des Équipements de Calcul Intensif en Fédération Wallonie Bruxelles (CÉCI) funded by the F.R.S.-FNRS under convention 2.5020.11. John A.~Lee is a Research Director with the F.R.S.-FNRS.

\section*{Conflic of interest statement}

Viktor Wase, Otte Marthin, Johan Sundstr\"om and Erik Engwall are employed at RaySearch Laboratories, Stockholm, Sweden. Guillaume Janssens and Kevin Souris are employed at Ion Beam Applications (IBA), Louvain-La-Neuve, Belgium. 


\clearpage

\bibliography{timeEff}      

\begin{thebibliography}{10}

\bibitem{otto_volumetric_2007}
Karl Otto.
\newblock Volumetric modulated arc therapy: {IMRT} in a single gantry arc: {Single} arc radiation therapy.
\newblock {\em Medical Physics}, 35(1):310--317, December 2007.

\bibitem{deasy_conformal_1995}
J.O. Deasy, T.R. Mackie, and P.M. DeLuca.
\newblock Conformal proton tomotherapy using distal-edge tracking.
\newblock {\em Radiotherapy and Oncology}, 37(3):S43, 1995.
\newblock Place: Ireland INIS Reference Number: 34044555.

\bibitem{sandison_phantom_1997}
G.A. Sandison, E.~Papiez, C.~Bloch, and J.~Morphis.
\newblock Phantom assessment of lung dose from proton arc therapy.
\newblock {\em International Journal of Radiation Oncology*Biology*Physics}, 38(4):891--897, July 1997.

\bibitem{pedroni_200-mev_1995}
Eros Pedroni, Reinhard Bacher, Hans Blattmann, Terence Böhringer, Adolf Coray, Antony Lomax, Shixiong Lin, Gudrun Munkel, Stefan Scheib, Uwe Schneider, and Alexander Tourovsky.
\newblock The 200-{MeV} proton therapy project at the {Paul} {Scherrer} {Institute}: {Conceptual} design and practical realization.
\newblock {\em Medical Physics}, 22(1):37--53, January 1995.

\bibitem{seco_proton_2013}
Joao Seco, Guan Gu, Tiago Marcelos, Hanne Kooy, and Henning Willers.
\newblock Proton {Arc} {Reduces} {Range} {Uncertainty} {Effects} and {Improves} {Conformality} {Compared} {With} {Photon} {Volumetric} {Modulated} {Arc} {Therapy} in {Stereotactic} {Body} {Radiation} {Therapy} for {Non}-{Small} {Cell} {Lung} {Cancer}.
\newblock {\em International Journal of Radiation Oncology*Biology*Physics}, 87(1):188--194, 2013.

\bibitem{ding_have_2018}
Xuanfeng Ding, Xiaoqiang Li, An~Qin, Jun Zhou, Di~Yan, Craig Stevens, Daniel Krauss, and Peyman Kabolizadeh.
\newblock Have we reached proton beam therapy dosimetric limitations? – {A} novel robust, delivery-efficient and continuous spot-scanning proton arc ({SPArc}) therapy is to improve the dosimetric outcome in treating prostate cancer.
\newblock {\em Acta Oncologica}, 57(3):435--437, March 2018.
\newblock Number: 3.

\bibitem{ding_improving_2019}
Xuanfeng Ding, Jun Zhou, Xiaoqiang Li, Kevin Blas, Gang Liu, Yinan Wang, An~Qin, Prakash Chinnaiyan, Di~Yan, Craig Stevens, Inga Grills, and Peyman Kabolizadeh.
\newblock Improving dosimetric outcome for hippocampus and cochlea sparing whole brain radiotherapy using spot-scanning proton arc therapy.
\newblock {\em Acta Oncologica}, 58(4):483--490, April 2019.
\newblock Number: 4.

\bibitem{li_improve_2018}
Xiaoqiang Li, Peyman Kabolizadeh, Di~Yan, An~Qin, Jun Zhou, Ye~Hong, Thomas Guerrero, Inga Grills, Craig Stevens, and Xuanfeng Ding.
\newblock Improve dosimetric outcome in stage {III} non-small-cell lung cancer treatment using spot-scanning proton arc ({SPArc}) therapy.
\newblock {\em Radiation Oncology}, 13(1):35, December 2018.

\bibitem{liu_improve_2020}
Gang Liu, Xiaoqiang Li, An~Qin, Weili Zheng, Di~Yan, Sheng Zhang, Craig Stevens, Peyman Kabolizadeh, and Xuanfeng Ding.
\newblock Improve the dosimetric outcome in bilateral head and neck cancer ({HNC}) treatment using spot-scanning proton arc ({SPArc}) therapy: a feasibility study.
\newblock {\em Radiation Oncology}, 15(1):21, December 2020.
\newblock Number: 1.

\bibitem{liu_novel_2020}
Gang Liu, Xiaoqiang Li, Lewei Zhao, Weili Zheng, An~Qin, Sheng Zhang, Craig Stevens, Di~Yan, Peyman Kabolizadeh, and Xuanfeng Ding.
\newblock A novel energy sequence optimization algorithm for efficient spot-scanning proton arc ({SPArc}) treatment delivery.
\newblock {\em Acta Oncologica}, 59(10):1178--1185, October 2020.
\newblock Number: 10.

\bibitem{de_jong_proton_2023}
Bas~A. De~Jong, Erik~W. Korevaar, Anneke Maring, Chimène~I. Werkman, Daniel Scandurra, Guillaume Janssens, Stefan Both, and Johannes~A. Langendijk.
\newblock Proton arc therapy increases the benefit of proton therapy for oropharyngeal cancer patients in the model based clinic.
\newblock {\em Radiotherapy and Oncology}, 184:109670, July 2023.

\bibitem{ding_spot-scanning_2016}
Xuanfeng Ding, Xiaoqiang Li, J.~Michele Zhang, Peyman Kabolizadeh, Craig Stevens, and Di~Yan.
\newblock Spot-{Scanning} {Proton} {Arc} ({SPArc}) {Therapy}: {The} {First} {Robust} and {Delivery}-{Efficient} {Spot}-{Scanning} {Proton} {Arc} {Therapy}.
\newblock {\em International Journal of Radiation Oncology*Biology*Physics}, 96(5):1107--1116, December 2016.
\newblock Number: 5.

\bibitem{battinelli_proton_2019}
Cecilia Battinelli.
\newblock Proton {Arc} {Therapy} {Optimization}.
\newblock {\em Master thesis}, 2019.

\bibitem{engwall_fast_2022}
Erik Engwall, Cecilia Battinelli, Viktor Wase, Otte Marthin, Lars Glimelius, Rasmus Bokrantz, Björn Andersson, and Albin Fredriksson.
\newblock Fast robust optimization of proton {PBS} arc therapy plans using early energy layer selection and spot assignment.
\newblock {\em Physics in Medicine \& Biology}, 67(6):065010, March 2022.
\newblock Number: 6.

\bibitem{zhang_energy_2022}
Gezhi Zhang, Haozheng Shen, Yuting Lin, Ronald~C Chen, Yong Long, and Hao Gao.
\newblock Energy layer optimization via energy matrix regularization for proton spot‐scanning arc therapy.
\newblock {\em Medical Physics}, 49(9):5752--5762, September 2022.
\newblock Number: 9.

\bibitem{wuyckens_treatment_2022}
Sophie Wuyckens, Michael Saint-Guillain, Guillaume Janssens, Lewei Zhao, Xiaoqiang Li, Xuanfeng Ding, Edmond Sterpin, John~A. Lee, and Kevin Souris.
\newblock Treatment planning in arc proton therapy: {Comparison} of several optimization problem statements and their corresponding solvers.
\newblock {\em Computers in Biology and Medicine}, 148:105609, September 2022.

\bibitem{teoh_volumetric_2011}
M~Teoh, C~H Clark, K~Wood, S~Whitaker, and A~Nisbet.
\newblock Volumetric modulated arc therapy: a review of current literature and clinical use in practice.
\newblock {\em The British Journal of Radiology}, 84(1007):967--996, November 2011.

\bibitem{chang_feasibility_2020}
Sheng Chang, Gang Liu, Lewei Zhao, Joshua~T. Dilworth, Weili Zheng, Saada Jawad, Di~Yan, Peter Chen, Craig Stevens, Peyman Kabolizadeh, Xiaoqiang Li, and Xuanfeng Ding.
\newblock Feasibility study: spot-scanning proton arc therapy ({SPArc}) for left-sided whole breast radiotherapy.
\newblock {\em Radiation Oncology}, 15(1):232, December 2020.
\newblock Number: 1.

\bibitem{van_de_water_shortening_2020}
Steven Van De~Water, Maria~F Belosi, Francesca Albertini, Carla Winterhalter, Damien~C Weber, and Antony~J Lomax.
\newblock Shortening delivery times for intensity-modulated proton therapy by reducing the number of proton spots: an experimental verification.
\newblock {\em Physics in Medicine \& Biology}, 65(9):095008, May 2020.
\newblock Number: 9.

\bibitem{krieger_quantitative_2022}
Miriam Krieger, Steven Van De~Water, Michael~M. Folkerts, Alejandro Mazal, Silvia Fabiano, Nicola Bizzocchi, Damien~C. Weber, Sairos Safai, and Antony~J. Lomax.
\newblock A quantitative {FLASH} effectiveness model to reveal potentials and pitfalls of high dose rate proton therapy.
\newblock {\em Medical Physics}, 49(3):2026--2038, March 2022.

\bibitem{wera_proton_2023}
Anne-Catherine Wéra, Macarena Chocan~Vera, Hamdiye Ozan, Erik Engwall, Viktor Wase, Otte Marthin, Johan Sundström, Sophie Wuyckens, Karine Haustermans, Ana~M. Barragán‐Montero, Kevin Souris, John~A. Lee, and Edmond Sterpin.
\newblock Proton arc therapy for esophageal cancer: is the dosimetric benefit worth the effort?
\newblock {\em Submitted to Physica Medica}, 2023.

\bibitem{de_jong_spot_2023}
Bas~A. De~Jong, Cecilia Battinelli, Jeffrey Free, Dirk Wagenaar, Erik Engwall, Guillaume Janssens, Johannes~A. Langendijk, Erik~W. Korevaar, and Stefan Both.
\newblock Spot scanning proton arc therapy reduces toxicity in oropharyngeal cancer patients.
\newblock {\em Medical Physics}, 50(3):1305--1317, March 2023.
\newblock Number: 3.

\bibitem{wase_optimizing_2024}
Viktor Wase, Otte Marthin, Albin Fredriksson, and Anton Finnson.
\newblock Optimizing the traversal time for gantry trajectories for proton arc therapy treatment plans.
\newblock {\em Physics in Medicine \& Biology}, February 2024.

\bibitem{gu_novel_2020}
Wenbo Gu, Dan Ruan, Qihui Lyu, Wei Zou, Lei Dong, and Ke~Sheng.
\newblock A novel energy layer optimization framework for spot‐scanning proton arc therapy.
\newblock {\em Medical Physics}, 47(5):2072--2084, May 2020.
\newblock Number: 5.

\bibitem{zhao_first_2023}
Lewei Zhao, Juntao You, Gang Liu, Sophie Wuyckens, Xiliang Lu, and Xuanfeng Ding.
\newblock The first direct method of spot sparsity optimization for proton arc therapy.
\newblock {\em Acta Oncologica}, 62(1):48--52, January 2023.

\bibitem{wase_proton_2024}
Viktor Wase, Sophie Wuyckens, John~A. Lee, and Michael Saint-Guillain.
\newblock The proton arc therapy treatment planning problem is {NP}-{Hard}.
\newblock {\em Computers in Biology and Medicine}, 171:108139, March 2024.

\bibitem{cao_intensity_2023}
Wenhua Cao, Yupeng Li, Xiaodong Zhang, Falk Poenisch, Pablo Yepes, Narayan Sahoo, David Grosshans, Susan McGovern, G.~Brandon Gunn, Steven~J. Frank, and Xiaorong~R. Zhu.
\newblock Intensity modulated proton arc therapy via geometry‐based energy selection for ependymoma.
\newblock {\em Journal of Applied Clinical Medical Physics}, 24(7):e13954, July 2023.

\bibitem{zhang_treatment_2023}
Gezhi Zhang, Yong Long, Yuting Lin, Ronald~C Chen, and Hao Gao.
\newblock A treatment plan optimization method with direct minimization of number of energy jumps for proton arc therapy.
\newblock {\em Physics in Medicine \& Biology}, 68(8):085001, April 2023.
\newblock Number: 8.

\bibitem{craft_exploration_2005}
David Craft, Tarek Halabi, and Thomas Bortfeld.
\newblock Exploration of tradeoffs in intensity-modulated radiotherapy.
\newblock {\em Physics in Medicine and Biology}, 50(24):5857--5868, December 2005.

\bibitem{wuyckens_bi-criteria_2022}
Sophie Wuyckens, Lewei Zhao, Michael Saint-Guillain, Guillaume Janssens, Edmond Sterpin, Kevin Souris, Xuanfeng Ding, and John~A Lee.
\newblock Bi-criteria {Pareto} optimization to balance irradiation time and dosimetric objectives in proton arc therapy.
\newblock {\em Physics in Medicine \& Biology}, 67(24):245017, December 2022.

\bibitem{fu_simultaneous_2024}
Anqi Fu, Vicki~T. Taasti, and Masoud Zarepisheh.
\newblock Simultaneous reduction of number of spots and energy layers in intensity modulated proton therapy for rapid spot scanning delivery.
\newblock {\em Medical Physics}, 51(8):5722--5737, August 2024.

\bibitem{langendijk_indicatie_2019}
JA~Langendijk, MS~Hoogeman, R~Monshouwer, and M~Verheij.
\newblock Indicatie {Protocol} {Protonen} {Therapie} (versie 2.2) {HOOFD}-{HALSTU}- {MOREN}.
\newblock Technical report, Landelijk Platform Protonentherapi, 2019.

\end{thebibliography}
\bibliographystyle{unsrt}
\clearpage
\section*{Supplementary Material}
\beginsupplement

\section*{SM1. Filtering algorihtm}

Algorithm \ref{algo6:filtering} shows the filtering workflow followed in the study. Each element $s_i$ from the MaxBoundSequence $\bm{s}$, is the maximum number of EL or SU we want to achieve after the filtering in the run $i$.

\begin{algorithm}

\caption{Filtering algorithm workflow}\label{alg6:cap}
\hspace*{\algorithmicindent} \textbf{Input data:} \text{Optimized spot weight} $\bm{x}^*$, $N$ EL, $M$ SU, MaxBoundSequence $\bm{s}$, $R$ runs, filtering algorithm $\mathcal{A}$
\begin{algorithmic}[1]
\Require $\{ s_i\}_{i=1}^{R}: s_i \geq s_{i+1} \forall i \in \mathbb{N}$ \Comment{Monotonic decreasing sequence}
\Ensure $\bm{x} = \bm{x}^*$
\For{ i in range(R)} 
\If{$\mathcal{A}$ is unrestricted or SU gap}
\State Filter $N-s_i$ EL 
\ElsIf{$\mathcal{A}$ is SU}
\State Filter $M-s_i$ SU
\EndIf
\For{ i in range(n\_iter)}
\State $\bm{x} \gets \text{spot\_weights\_optimizer}(\bm{x})$ \Comment{Continue optimization}
\EndFor
\State save\_spot\_weights($\bm{x}$) \Comment{Save plan properties}
\State $\bm{x} \gets \bm{x}^*$ \Comment{Reset to initial optimized spot weights}
\EndFor

\end{algorithmic}
\label{algo6:filtering}
\end{algorithm}

\section*{SM2. IMPT and PAT treatment planning}
\subsection*{A. Oropharyngeal cohort}

With the four selected patients being bilateral cases, the proton plans were designed using a simultaneous integrated boost IMPT technique which treats different target volumes to different dose levels within a single treatment session without increasing the toxicity. For all plans, we use a constant relative biological effectiveness (RBE) factor of 1.1. The prescribed dose was 70 Gy to the high-risk clinical target volume (CTV) and 54.25 Gy to the low-risk CTV nodal regions, delivered in 35 fractions. The plan was designed with four beams at couch and gantry angles of (10°, 60°), (10°, 120°), (350°, 240°), and (350°, 300°), respectively. A range shifter was systematically used for each beam. Monte Carlo dose calculation and robust worst-case optimization were employed for the proton plans, considering 21 scenarios (2.6\% proton range error and 4 mm systematic error in all three spatial directions) for the CTV volume, the spinal cord and the brainstem.

\begin{table}[!ht]
\centering
\resizebox{\textwidth}{!}{%
\begin{tabular}{lll}
\toprule
\textbf{Region of interest    }          & \textbf{Clinical goal  }                                      & \textbf{Worst case }                                       \\ \midrule
Low-risk CTV*                   & D98\% \textgreater{}= 51.54 Gy (95\% Dp = 54.25 Gy)) & D98\% \textgreater{}= 53.17 Gy (98\% Dp = 54.25 Gy)  \\
\multirow{2}{*}{High-risk CTV*} & D98\% \textgreater{}= 66.50 Gy (95\% Dp = 70 Gy))    & D98\% \textgreater{}= 68.60 Gy (98\% Dp = 70 Gy)) \\
                                & D0.03cc \textgreater ~74.90 Gy (107\% Dp = 70Gy)      &                                                   \\
Body                            & D0.03cc \textless ~74.90 Gy (107\% Dp = 70Gy)         &                                                   \\
Spinal cord                     & D0.03cc \textless ~45 Gy                              &                                                   \\
Brain stem                      & D0.03cc \textless ~54 Gy                              &                                                   \\
PCMSup/PCMMid/PCMInf            & Dmean \textless ~40 Gy                                &                                                   \\
Parotids (left and right)       & Dmean \textless ~26 Gy                                &                                                   \\
SMGs (left and right)           & Dmean \textless  ~40 Gy                               &                                                   \\
Oral cavity                     & Dmean \textless ~ 35 Gy                               &                                                   \\
Larynx                          & Dmean \textless  ~50 Gy                               &                                                   \\
Mandible                        & D0.03cc \textless ~70 Gy                              &                                                   \\
Glottic area                    & Dmean \textless  ~40 Gy                               &                                                   \\
Esophagus                       & Dmean \textless  ~35 Gy                               &                                                   \\ 
Choclea (left and right)        & Dmean \textless  ~30 Gy                               &                                                  \\ \bottomrule
\end{tabular}
}
\caption{Clinical goals for IMPT and PAT oropharyngeal cancer planning. Different criteria were applied for nominal and worst case scenarios. (*) means the clinical goal was included as a robust objective during the optimization process. bbreviations: PCM = pharyngeal constrictor muscles, Sup = Superior, Mid = Mid = Middle, Inf = Inferior, SMG:
submandibular glands }
\label{tab:hanGoals}
\end{table}

The PAT plans were designed with a single full arc beam with one-degree spacing between each control point, and delivered clockwise. This results in 360 initial energy layers available for optimization. No range shifter was employed in designing the arc plans, as we assumed that the use of multiple directions and energy layer selection could alleviate the necessity for a range shifter, while also preventing the widening of the penumbra as pointed out by Ding et al (Journal of Applied Clinical Medical Physics, 2018). They were planned using a total of 250 iterations and spot filtration iteration set at 150. The Monte Carlo (MC) dose calculation was carried out until it reached an uncertainty of 0.5\%. Clinical goals used for treatment plan evaluation are listed in Table \ref{tab:hanGoals}.

\subsection*{B. Lung cohort}

The retrospective analysis was conducted on 4 patients with unresectable lung cancer selected out of a database of 14 patients. For each patient, a 4D-CT scan containing 10 phases was available. Gross tumor volume (GTV) was delineated on the mid-position image (MidP CT) while the clinical tumor volume (CTV) was created as an isotropic 5 mm expansion of the GTV. Patient-specific IMPT plans had been previously generated for all 14 patients to be used in a prior study (Badiu et al, Phys. Med., 2016). All IMPT plans contained 3 beams, placed  according to tumor extension and location, to avoid the spinal canal and spare contralateral lung as much as possible. PAT plans were created using one partial arc per plan, and only one revolution around the patient (see table \ref{tab:lungBeams}). Gantry spacing was set to 1 or 2 degrees, depending on the beam range and amount of total energy layers desired. No range shifter was needed for any patient

\begin{table}[h]
\centering
\begin{tabular}{lllllll}
\toprule
\textbf{Lung patients} & \multicolumn{3}{l}{\textbf{IMPT}}                                                                  & \multicolumn{3}{l}{\textbf{PAT}}                                   \\ \midrule
             & Gantry angles (°) & Couch angle & \begin{tabular}[c]{@{}l@{}}Range\\ shifter\end{tabular} & Arc start, stop (°) & Gantry angle spacing (°) & Rotation \\ \midrule
1            & (190,240,290)     & 0           & N                                                       & (310,190)           & 1                        & CCW      \\
2            & (200,270,340)     & 0           & N                                                       & (330,190)           & 1                        & CCW      \\
3            & (190,240,340)     & 0           & N                                                       & (0,180)             & 2                        & CCW      \\
4            &  (180,270,340)                  & 0           & N                                                       & (180,0)             & 1                        & CW      \\ \bottomrule
\end{tabular}
\caption{Plan characteristics: beam angles (IMPT), couch rotation, presence of a range shifter, arc range (PAT plans) and gantry angle spacing. Abbreviations: CW = clockwise, CCW = counter-clockwise.}
\label{tab:lungBeams}
\end{table}

4D worst-case minimax robust optimization was performed on a non isotropic CTV expansion, according to Van Herk’s formula. Monte Carlo dose calculations were employed with a prescription dose of 60 Gy in 30 fractions, normalized to D50\% = 60 Gy using a dose calculation grid size of 2.5 x 2.5 x 2.5 mm$^3$. The non-isotropic setup error was divided into two parts following the study by Badiu et al.: 5 mm isotropic error was factored into RayStation's setup parameters for robust optimization, while the remaining error was managed through non-isotropic CTV expansion. Both treatment modalities adhered to the same optimization criteria and clinical goals, balancing OAR sparing and target coverage (see Table \ref{tab:lungGoals}). However, optimization functions were customized for each modality. Robust objective functions were set on CTV and spinal canal. The robust optimization encompassed 84 scenarios: 7 (setup errors:  ±5 mm in x,y,z directions, plus the nominal scenario) × 3 (image conversion errors: ±3\%, 0\%) × 4 (breathing phases: MidP, maximum inhale, maximum exhale and an additional mid-ventilation phase).

\begin{table}[!ht]
\centering
\begin{tabular}{lll}
\toprule
\textbf{Region of interet}               & \textbf{Clinical goal}                & \textbf{Worst case }                 \\ \midrule
\multirow{2}{*}{CTV(*)} & D98\% \textgreater{}= 57 Gy & D95\% \textgreater{}= 57 Gy \\
                        & D1\% \textless~ 63 Gy        &                             \\
Esophagus               & D0.04cc \textless ~60 Gy     &                             \\
Heart                   & Dmean \textless ~20 Gy       &                             \\
                        & D0.04cc \textless ~60 Gy     &                             \\
Lungs - GTV             & Dmean \textless ~20 Gy       &                             \\
                        & V30Gy \textless ~20\%        &                             \\
Spinal canal(*)         & D0.04cc \textless ~50 Gy     & D0.04cc \textless ~50 Gy    \\
Body                    & D1cc \textless ~63 Gy        & D1cc \textless ~63 Gy     \\ \bottomrule  
\end{tabular}%
\caption{Clinical goals for IMPT and PAT lung cancer planning. Different criteria were applied for nominal and worst case scenarios. (*) means the clinical goal was included as a robust objective during the optimization process.}
\label{tab:lungGoals}
\end{table}

\section*{SM3. Machine delivery assumptions}

Table \ref{tab6:machineParam} details the machine delivery assumptions used by the ATOM algorithm to determine the gantry velocity curve. These selections were made based on realistic assumptions rather than referring to a particular proton machine's specifications. For example, it is acknowledged that, at present, switch-up times are typically an order of magnitude higher than switch-down times in most proton centers due to the magnetic hysteresis effect. Regarding the maximum gantry velocity, a slightly more conservative value was chosen, considering that current proton gantries typically rotate at one revolution per minute. The maximum window size specifies the angular safety tolerance around each irradiation direction. Note that the irradiation parameters referred in Table \ref{tab6:machineParam} were only used in the simple ATOM model. ATOM was also used with the ScanAlgo parameters from an IBA PPlus system, which are not listed here.

\begin{table}[!ht]
\centering
{\fontsize{10}{12}\selectfont
\begin{tabular}{lc} 
\toprule
\textbf{Maximum gantry velocity} & 5°/s \\
\textbf{Maximum gantry acceleration} & 0.5 °/s$^2$  \\
\textbf{Maximum gantry jerk} & 0.5 °/s$^3$ \\
\textbf{Downward energy switching time} & 0.5 s      \\ 
\textbf{Upward energy switching time} & 5 s        \\
\textbf{Spot switching time} & 2 ms       \\
\textbf{Spot delivery time (per MU)} & 5 ms \\
\textbf{Maximum window size}         & 0.99° \\ \bottomrule
\end{tabular}%
}
\caption{Machine delivery assumptions used for ATOM BDT estimation.}
\label{tab6:machineParam}
\end{table}
\newpage

\section*{SM4. Supplementary results}

\subsection*{A. OPC results}

\begin{longtable}{lllllll}
\toprule
\textbf{Patient}                & \multicolumn{6}{l}{\textbf{OPC1}}                                                    \\
\endfirsthead
\endhead
\textbf{Plan}                   & IMPT    & PAT    & Filtered PAT & IMPT       & PAT       & Filtered PAT \\
\textbf{Scenario}               & Nominal & Nominal & Nominal       & Worst case & Worst case & Worst case   \\ \midrule
\textbf{CTV70 D98}              & 68.76   & 69.13   & 69.1          & 68.29      & 68.95      & 68.84         \\
\textbf{CTV54.25 D98}           & 53.99   & 54.06   & 53.94         & 53.31      & 53.63      & 53.24         \\ \midrule
\textbf{Scenario}              & Nominal & Nominal & Nominal       & Mean ± std & Mean ± std & Mean ± std    \\ \midrule
\textbf{SpinalCord D0.03cm3}    & 30.13   & 32.54   & 32.47         & 33.85 ± 6.31 & 33.94 ± 3.94 & 34.07 ± 3.85   \\
\textbf{Brainstem D0.03cm3}     & 9.83    & 10.22   & 10.31         & 11.06 ± 3.68   & 11.95 ± 2.18 & 11.94 ± 2.16\\
\textbf{Parotid\_L Dmean}       & 14.68   & 12.92   & 12.87         &  15.01 ± 1.98    &13.22 ± 2.50&  13.16 ± 2.49\\
\textbf{Parotid\_R Dmean}       & 16.40    & 13.69   & 13.70          &  16.63 ± 2.36    &13.89 ± 2.45& 13.89 ± 2.46\\
\textbf{Submandibular\_L Dmean} & 64.66   & 65.96   & 66.05         &  64.69 ± 0.88   &65.93 ± 1.00& 66.04 ± 0.87\\
\textbf{Submandibular\_R Dmean} & 68.16   & 66.39   & 66.13         &67.88 ± 0.61&66.25 ± 0.68&  66.00 ± 0.62      \\
\textbf{Oral cavity Dmean}      & 54.49   & 54.91   & 54.73         &54.29 ± 2.49&54.77 ± 1.97&  54.59 ± 1.98      \\
\textbf{PCMSup Dmean  }         & 68.22   & 67.63   & 67.60          &67.98 ± 0.89&67.61 ± 0.84&   67.54 ± 0.84     \\
\textbf{PCMMid Dmean}           & 68.39   & 67.92   & 67.72         &68.01 ± 0.28&68.01 ± 0.28&    68.01 ± 0.28    \\
\textbf{PCMInf Dmean }          & 43.23   & 33.6    & 33.54         &43.45 ± 1.61&33.82 ± 1.98&   33.73 ± 1.99     \\
\textbf{Esophagus Dmean }       & 5.89    & 2.48    & 2.44          &6.09 ± 0.98&2.70 ± 0.63&   2.63 ± 0.62      \\ \midrule
\textbf{Patient}                & \multicolumn{6}{l}{\textbf{OPC2}}                                                    \\
\textbf{Plan  }                 & IMPT    & PAT    & Filtered PAT & IMPT       & PAT       & Filtered PAT \\
\textbf{Scenario }              & Nominal & Nominal & Nominal       & Worst case & Worst case & Worst case    \\ \midrule
\textbf{CTV70 D98  }            & 69.06   & 69.17   & 68.98         & 68.74      & 68.92      & 68.61         \\
\textbf{CTV54.25 D98 }          & 53.86   & 54.29   & 53.77         & 53.38      & 53.65      & 53.05         \\ \midrule
\textbf{Scenario}              & Nominal & Nominal & Nominal       & Mean ± std & Mean ± std & Mean ± std    \\ \midrule
\textbf{SpinalCord D0.03cm3}    & 29.68   & 7.15    & 6.99          &30.36 ± 4.81 & 8.04 ± 0.99&  7.96 ± 0.99     \\
\textbf{Brainstem D0.03cm3 }    & 28.95   & 2.67    & 2.64          &28.91 ± 6.37 & 3.04 ± 1.04 &   3.42 ± 1.06      \\
\textbf{Parotid\_L Dmean }      & 25.40    & 19.03   & 18.83         & 25.64 ± 2.23 & 19.29 ± 2.53 &  19.09 ± 2.50        \\
\textbf{Parotid\_R Dmean }      & 19.89   & 17.8    & 17.56         & 20.24 ± 2.16 & 18.07 ± 2.73 &  17.83 ± 2.67      \\
\textbf{Submandibular\_L Dmean} & 69.17   & 67.95   & 67.76         &69.04 ± 0.35 & 67.87 ± 0.45 &  67.75 ± 0.42      \\
\textbf{Submandibular\_R Dmean} & 69.65   & 68.21   & 67.89         &69.49 ± 0.29 & 68.13 ± 0.44 &  67.81 ± 0.38      \\
\textbf{Oral cavity Dmean}      & 49.97   & 50.02   & 49.92         &49.79 ± 2.88 & 49.94 ± 2.59 &  49.85 ± 2.57     \\
\textbf{PCMSup Dmean}           & 69.63   & 69.33   & 69.35         &69.49 ± 0.32 &69.21 ± 0.45 &  69.22 ± 0.43      \\
\textbf{PCMMid Dmean}           & 67.56   & 65.93   & 65.48         &66.32 ± 0.89 & 66.32 ± 0.89 &  66.32 ± 0.89      \\
\textbf{PCMInf Dmean  }         & 47.44   & 34.03   & 33.71         &48.03 ± 1.54 &34.94 ± 2.07 &  34.59 ± 2.07      \\
\textbf{Esophagus Dmean}        & 32.51   & 30.34   & 29.9          &32.91 ± 1.56 & 30.80 ± 2.01 &   30.34 ± 2.17     \\ \midrule
\textbf{Patient }               & \multicolumn{6}{l}{\textbf{OPC3}}                                                    \\
\textbf{Plan}                   & IMPT    & PAT    & Filtered PAT & IMPT       & PAT       & Filtered PAT \\
\textbf{Scenario }              & Nominal & Nominal & Nominal       & Worst case & Worst case & Worst case    \\ \midrule
\textbf{CTV70 D98 }             & 69.02   & 69.05   & 69.02         & 68.36      & 68.71      & 68.83         \\
\textbf{CTV54.25 D98 }          & 53.84   & 53.98   & 53.94         & 53.24      & 53.38      & 53.36         \\ \midrule
\textbf{Scenario}              & Nominal & Nominal & Nominal       & Mean ± std & Mean ± std & Mean ± std    \\ \midrule
\textbf{SpinalCord D0.03cm3 }   & 16.46   & 11.42   & 11.39         & 16.77 ± 4.83 & 14.30 ± 2.69 & 14.45 ± 2.61        \\
\textbf{Brainstem D0.03cm3 }    & 21.83   & 8.41    & 8.76          & 24.90 ± 6.57 & 11.25 ± 3.54 & 11.19 ± 3.56       \\
\textbf{Parotid\_L Dmean  }     & 34.29   & 26.81   & 26.83         & 34.38 ± 3.00& 26.96 ± 3.39& 26.97 ± 3.37        \\
\textbf{Parotid\_R Dmean }      & 32.42   & 26.05   & 25.88         & 32.53 ± 2.61& 26.24 ± 3.10& 26.08 ± 3.10      \\
\textbf{Submandibular\_L Dmean} & 69.73   & 69.74   & 69.63         & 69.65 ± 0.41& 69.75 ± 0.13& 69.66 ± 0.11       \\
\textbf{Submandibular\_R Dmean} & 68.96   & 68.64   & 68.58         & 68.72 ± 0.72& 68.45 ± 0.46& 68.39 ± 0.6       \\
\textbf{Oral cavity Dmean }     & 39.45   & 37.48   & 37.42         & 39.47 ± 2.38& 37.56 ± 2.47& 37.49 ± 2.48       \\
\textbf{PCMSup Dmean}           & 61.70    & 59.48   & 59.43         & 61.59 ± 1.12& 59.58 ± 1.51&  59.52 ± 1.51      \\
\textbf{PCMMid Dmean}           & 58.79   & 57.67   & 57.69         & 58.05 ± 0.52& 58.05 ± 0.52& 58.05 ± 0.52      \\
\textbf{PCMInf Dmean}           & 4.94    & 7.51    & 7.43          & 5.39 ± 0.82& 8.77 ± 1.12&  8.71 ± 1.14       \\
\textbf{Esophagus Dmean}        & 4.22    & 3.04    & 3.03          & 4.64 ± 1.37& 3.57 ± 1.04& 3.55 ± 1.04        \\ \midrule
\textbf{Patient}                & \multicolumn{6}{l}{\textbf{OPC4}}                                                    \\
\textbf{Plan }                  & IMPT    & PAT    & Filtered PAT & IMPT       & PAT       & Filtered PAT \\
\textbf{Scenario}               & Nominal & Nominal & Nominal       & Worst case & Worst case & Worst case    \\ \midrule
\textbf{CTV70 D98 }             & 68.12   & 69.12   & 69.05         & 67.57      & 68.71      & 68.7          \\
\textbf{CTV54.25 D98 }          & 53.56   & 54.58   & 54.47         & 52.83      & 53.87      & 53.85         \\ \midrule
\textbf{Scenario}              & Nominal & Nominal & Nominal       & Mean ± std & Mean ± std & Mean ± std    \\ \midrule
\textbf{SpinalCord D0.03cm3}    & 14.95   & 20.55   & 20.72         & 16.05 ± 5.63 & 24.04 ± 2.62 & 24.17 ± 2.69       \\
\textbf{Brainstem D0.03cm3 }    & 10.49   & 11.04   & 10.89         & 11.67 ± 4.61 & 11.96 ± 4.00 & 12.08 ± 4.07       \\
\textbf{Parotid\_L Dmean}       & 26.91   & 26.41   & 26.46         & 27.07 ± 2.99 & 26.43 ± 2.42 & 26.47 ± 2.40       \\
\textbf{Parotid\_R Dmean}       & 29.09   & 26.49   & 26.41         & 29.21 ± 2.47 & 26.51 ± 2.39 & 26.44 ± 2.35      \\
\textbf{Submandibular\_L Dmean} & 43.46   & 38.82   & 38.42         & 43.19 ± 3.72 & 38.80 ± 2.73 & 38.41 ± 2.79      \\
\textbf{Submandibular\_R Dmean} & 57.95   & 52.13   & 52.13         & 57.40 ± 3.24 & 52.15 ± 3.40 & 52.14 ± 3.38     \\
\textbf{Oral cavity Dmean}      & 18.31   & 17.09   & 16.99         & 18.26 ± 2.09 & 17.19 ± 1.99 & 17.09 ± 1.98     \\
\textbf{PCMSup Dmean}           & 60.10    & 58.20    & 58.15         & 59.83 ± 1.18& 58.31 ± 0.89 & 58.27 ± 0.95      \\
\textbf{PCMMid Dmean}           & 63.64   & 62.2    & 62.25         & 62.70 ± 0.67 & 62.70 ± 0.67 & 62.70 ± 0.67     \\
\textbf{PCMInf Dmean}           & 39.57   & 39.41   & 39.56         & 39.46 ± 3.50 & 39.83 ± 2.02 & 39.95 ± 2.03       \\
\textbf{Esophagus Dmean}        & 27.03   & 25.32   & 25.37         & 27.73 ± 4.32 & 27.93 ± 2.71 & 27.94 ± 2.79   \\ \bottomrule
\caption{Dose metrics for oropharyngeal cancer (OPC) patients reported for IMPT, PAT and filtered PAT plans in nominal and worst case scenarios.}
\label{tab:my-table}\\
\end{longtable}

\subsection*{B. LUNG results}

\begin{longtable}{lllllll}
\toprule
\textbf{Patient}               & \multicolumn{6}{l}{\textbf{LUNG1}}                                                   \\
\endfirsthead
\endhead
\textbf{Plan}                  & IMPT    & PAT    & Filtered PAT & IMPT       & PAT       & Filtered PAT \\
\textbf{Scenario}              & Nominal & Nominal & Nominal       & Worst case & Worst case & Worst case    \\ \midrule
\textbf{CTV D98}               & 58.90    & 58.96   & 59.08         &    55.22        &   57.00         &      57.25         \\
\textbf{CTV D95}               & 59.13   & 59.19   & 59.31         &     57.01       &     57.93       &       57.96        \\
\textbf{CTV D1 }               & 61.20    & 61.62   & 61.67         &     61.80       &     64.94       &       64.96        \\ \midrule
\textbf{Scenario}              & Nominal & Nominal & Nominal       & Mean ± std & Mean ± std & Mean ± std    \\ \midrule
\textbf{Esophagus D0.04cm3 }   & 26.65   & 23.83   & 24.20          &  28.17 ±  6.75       &  27.32 ± 7.05        &  27.69 ± 7.08           \\
\textbf{Lungs-GTV Dmean}       & 7.63    & 7.88    & 7.63          &   7.71 ±  0.12       &  7.71 ± 0.12       &    7.71 ± 0.12         \\
\textbf{Lungs-GTV V30}         & 8.31    & 9.08    & 8.43          &  8.58 ±  1.52       &   9.50 ± 1.76       &   8.86 ± 1.69          \\
\textbf{Heart Dmean}           & 4.87    & 5.09    & 5.09          &  4.62 ±  0.96       &  4.92 ± 1.04       &   4.92 ± 1.03           \\
\textbf{Heart D0.04cm3 }       & 62.47   & 63.11   & 63.11         &  62.49 ±  0.47       &   67.48 ± 2.46       &   67.31 ± 2.43           \\
\textbf{Spinal Canal D0.04cm3} & 18.86   & 8.19    & 8.33          &  18.33 ±  0.94       &   8.99 ± 2.65       & 9.04 ± 2.67            \\
\textbf{Body D1cm3 }           & 62.18   & 62.24   & 62.18         &   62.27 ± 0.16      &   65.45 ± 1.46       &   65.42 ± 1.45          \\ \midrule
\textbf{Patient}               & \multicolumn{6}{l}{\textbf{LUNG2}}                                                   \\
\textbf{Plan }                 & IMPT    & PAT    & Filtered PAT & IMPT       & PAT       & Filtered PAT \\
\textbf{Scenario}              & Nominal & Nominal & Nominal       & Worst case & Worst case & Worst case    \\ \midrule
\textbf{CTV D98}               & 58.36   & 58.64   & 58.71         & 55.98      & 57.57      & 57.58         \\
\textbf{CTV D95}               & 58.62   & 58.92   & 58.98         & 57.31      & 58.06      & 58.04         \\
\textbf{CTV D1}                & 61.88   & 61.44   & 61.47         & 63.09      & 63.12      & 63.01         \\ 
\textbf{Spinal Canal D0.04cm3} & 38.59   & 35.27   & 34.82         & 48.58 & 47.30 &   46.81     \\ \midrule
\textbf{Scenario}              & Nominal & Nominal & Nominal       & Mean ± std & Mean ± std & Mean ± std    \\ \midrule
\textbf{Esophagus D0.04cm3}    & 61.44   & 60.23   & 59.97         & 61.78 ± 0.73& 62.29 ± 0.90 & 62.20 ± 0.93    \\
\textbf{Lungs-GTV Dmean}       & 13.13   & 14.47   & 14.34         & 13.98 ± 0.60 & 13.98± 0.60 &    13.98 ± 0.6  \\
\textbf{Lungs-GTV V30}         & 16.08   & 16.09   & 15.90          & 16.05 ± 0.77 & 16.36 ± 1.06 &  16.17 ±  1.07   \\
\textbf{Heart Dmean }          & 6.02    & 7.02    & 6.99          & 5.67 ± 1.63 & 6.67 ± 1.74 &    6.64 ±  1.72   \\
\textbf{Heart D0.04cm3}        & 62.69   & 62.00    & 61.63         & 63.22 ± 1.32 & 62.49 ± 0.71 &   62.39 ± 0.86   \\
\textbf{Body D1cm3 }           & 63.95   & 62.60  & 62.52         & 64.14 ± 0.37 & 64.72 ± 2.20 &    64.65 ± 2.21  \\ \midrule
\textbf{Patient}               & \multicolumn{6}{l}{\textbf{LUNG3}}                                                   \\
\textbf{Plan}                  & IMPT    & PAT    & Filtered PAT & IMPT       & PAT       & Filtered PAT \\
\textbf{Scenario}              & Nominal & Nominal & Nominal       & Worst case & Worst case & Worst case    \\ \midrule
\textbf{CTV D98}               & 58.51   & 58.82   & 58.81         & 55.39      & 56.56      & 56.54         \\
\textbf{CTV D95}               & 58.77   & 59.08   & 59.05         & 57.02      & 57.57      & 57.56         \\
\textbf{CTV D1 }               & 61.73   & 61.50    & 61.57         & 62.97      & 64.60       & 64.66         \\
\textbf{Spinal Canal D0.04cm3} & 31.35   & 29.28   & 29.26         & 43.24 & 47.71 &  47.34      \\ \midrule
\textbf{Scenario}              & Nominal & Nominal & Nominal       & Mean ± std & Mean ± std & Mean ± std    \\ \midrule
\textbf{Esophagus D0.04cm3}    & 63.67   & 61.44   & 61.52         & 63.49 ± 0.96 & 63.79 ± 1.33& 63.85 ± 1.29    \\
\textbf{Lungs-GTV Dmean}       & 16.12   & 14.96   & 14.91         & 15.33 ± 0.56 &15.33 ± 0.56 &  15.33 ± 0.56  \\
\textbf{Lungs-GTV V30}         & 22.73   & 19.47   & 19.42         & 22.46 ± 2.53 &19.57 ± 1.86 &  19.51 ± 1.87  \\
\textbf{Heart Dmean}           & 9.80     & 9.05    & 9.06          & 9.78 ± 1.45 &9.09 ± 1.78&    9.10 ± 1.78  \\
\textbf{Heart D0.04cm3}        & 61.73   & 61.25   & 61.30          & 61.81 ± 0.50 &64.80 ± 2.77 &   64.92 ± 2.85   \\
\textbf{Body D1cm3}            & 63.76   & 62.14   & 62.02         & 63.99 ± 0.45 &65.78 ± 2.13&   65.79  ± 2.16   \\ \midrule
\textbf{Patient}               & \multicolumn{6}{l}{\textbf{LUNG4}}                                                   \\
\textbf{Plan}                  & IMPT    & PAT    & Filtered PAT & IMPT       & PAT       & Filtered PAT \\
\textbf{Scenario}              & Nominal & Nominal & Nominal       & Worst case & Worst case & Worst case    \\ \midrule
\textbf{CTV D98}               & 58.54   & 58.52   & 58.53         & 55.44      & 56.21      & 56.25         \\
\textbf{CTV D95}               & 58.83   & 58.84   & 58.82         & 57.05      & 57.07      & 57.15         \\
\textbf{CTV D1}                & 61.78   & 62.20    & 62.19         & 64.77      & 63.76      & 63.62         \\ 
\textbf{Spinal Canal D0.04cm3} & 31.35   & 29.28   & 29.26         & 49.35 & 45.35 & 45.54       \\ \midrule
\textbf{Scenario}              & Nominal & Nominal & Nominal       & Mean ± std & Mean ± std & Mean ± std    \\ \midrule
\textbf{Esophagus D0.04cm3 }   & 61.08   & 60.98   & 61.20          &   62.60 ± 1.38  &  62.60 ± 1.08   &   62.55 ± 1.14    \\
\textbf{Lungs-GTV Dmean}       & 14.61   & 9.46    & 9.40           &   11.16 ± 2.44  &  11.16 ± 2.44   &    11.16 ± 2.44    \\
\textbf{Lungs-GTV V30}         & 21.64   & 11.31   & 11.24         &   21.41 ± 3.43   &  11.45 ±  2.02  &    11.38 ± 2.02    \\
\textbf{Heart Dmean  }         & 9.40     & 7.33    & 7.27          &   9.10 ± 1.20  &   7.1 ± 0.70   &    7.05 ± 0.69    \\
\textbf{Heart D0.04cm3 }       & 63.44   & 63.44   & 63.44         &  63.94 ± 1.27  &   63.23 ± 1.2 &    63.15 ± 1.17   \\
\textbf{Body D1cm3 }           & 63.58   & 63.66   & 63.39         &   64.34 ± 0.85  &   64.98 ± 1.05  &    64.8 ± 1.02    \\ \bottomrule
\caption{Dose metrics for lung patients reported for IMPT, PAT and filtered PAT plans in nominal and worst case scenarios.}
\label{tab:lungResults}\\
\end{longtable}

\end{document}